\documentclass[twocolumn]{aastex631}

\usepackage{amsmath} 
\usepackage{booktabs}
\usepackage{multirow}
\usepackage{graphicx}
\usepackage{placeins}
\usepackage{enumitem}

\shorttitle{WASP-76b}
\shortauthors{Ramos Rosado et al.}

\begin{document}

\title{Detection of escaping magnesium in the upper atmosphere of WASP-76b}

\author[0000-0002-1056-3144]{Lakeisha M. Ramos Rosado}
\affiliation{William H. Miller III Department of Physics and Astronomy, Johns Hopkins University, 3400 N. Charles Street, Baltimore, MD 21218, USA}
\correspondingauthor{Lakeisha M. Ramos Rosado}
\email{lramosr1@jhu.edu}

\author[0000-0002-2248-3838]{Leonardo A. Dos Santos}
\affiliation{Space Telescope Science Institute, 3700 San Martin Drive, Baltimore, MD 21218, USA}
\affiliation{William H. Miller III Department of Physics and Astronomy, Johns Hopkins University, 3400 N. Charles Street, Baltimore, MD 21218, USA}

\author[0000-0003-3667-8633]{Joshua D. Lothringer}
\affiliation{Space Telescope Science Institute, 3700 San Martin Drive, Baltimore, MD 21218, USA}

\author[0000-0001-6050-7645]{David K. Sing}
\affiliation{William H. Miller III Department of Physics and Astronomy, Johns Hopkins University, 3400 N. Charles Street, Baltimore, MD 21218, USA}
\affiliation{Department of Earth and Planetary Sciences, Johns Hopkins University, 3400 N. Charles Street, Baltimore, MD 21218, USA}

\author[0009-0008-2801-5040]{Johanna Teske}
\affiliation{Earth and Planets Laboratory, Carnegie Institution for Science, 5241 Broad Branch Road, NW, Washington, DC 20015, USA}
\affiliation{The Observatories of the Carnegie Institution for Science, 813 Santa Barbara St., Pasadena, CA 91101, USA}

\author[0000-0001-5371-2675]{Aline A. Vidotto}
\affiliation{Leiden Observatory, Leiden University, PO Box 9513, 2300 RA Leiden, The Netherlands}

\author[0000-0001-9704-5405]{David Ehrenreich}
\affiliation{Observatoire astronomique de l’Université de Genève, chemin Pegasi 51, 1290 Versoix, Switzerland}

\author[0000-0002-7990-9596]{Julia V. Seidel}
\affiliation{European Southern Observatory, Santiago, Chile}
\affiliation{Université Côte d’Azur, Observatoire de la Côte d’Azur, CNRS, Laboratoire Lagrange, Nice, France}

\author[0000-0001-8981-6759]{H. Jens Hoeijmakers}
\affiliation{Lund Observatory, Division of Astrophysics, Department of Physics, Lund University, Box 118, 221 00 Lund, Sweden}

\author[0000-0003-4157-832X]{Munazza K. Alam}
\affiliation{Space Telescope Science Institute, 3700 San Martin Drive, Baltimore, MD 21218, USA}

\author[0000-0001-9699-1459]{Monika Lendl}
\affiliation{Observatoire astronomique de l’Université de Genève, 51 chemin Pegasi, 1290 Sauverny, Switzerland}

\begin{abstract}

Near-ultraviolet (NUV) observations can uniquely probe low‑density, high‑altitude gas in ultra‑hot Jupiters (UHJs) via strong resonance lines of metals. WASP‑76b is a well‑studied UHJ with prior detections of refractory metals and complex day–night chemistry, making it a prime target for atmospheric escape diagnostics in the NUV. We present HST/STIS E230M (2274 $-$ 3119 {\AA}) transmission spectroscopy of WASP-76b. While broad- ($\sim$ 200 {\AA}) and narrow-band ({4 \AA}) transmission spectra do not show spectral features, a velocity-resolved single line analysis reveals clear excess absorption in both components of the Mg\,{\sc ii} doublet (2796 {\AA} and 2803 {\AA}, $k$ and $h$ lines). The Mg\,{\sc ii} signal is localized to the blue wing (Doppler velocities [-50, -10] km s$^{-1}$) and is recovered with narrower 0.5 {\AA} bins; the inferred absorption exceeds both the optical continuum radius and the Roche-lobe scale, suggesting a highly extended upper atmosphere. We estimate a mass-loss rate of $\dot{M}\sim10^{12.14}\,\mathrm{g\,s^{-1}}$ from ATES. The present NUV data do not improve constraints on lower-atmosphere properties, and a comparison with upper-atmosphere forward models disfavor scenarios with extreme Mg enrichment. Together with the near-solar metallicity inferred from optical+IR spectra, this suggests that the Mg\,{\sc ii} absorption can arise in an escaping upper atmosphere without requiring extreme Mg enrichment, qualitatively consistent with hydrodynamic escape. Ongoing HST programs expanding the sample of UHJs focused on NUV observations will enable comparative tests of composition and kinematics in escaping atmospheres.

\end{abstract}


\section{Introduction} \label{sec:Intro}

Among the most remarkable exoplanet discoveries are planets on extremely close-in orbits, where intense stellar X-ray and extreme-UV (EUV) irradiation deposits energy in the upper atmosphere. This heating can inflate the atmosphere and, when the energy input is sufficient, drive hydrodynamic outflows (potentially reaching supersonic speeds) and atmospheric escape \citep{Watson_1981,Yelle_2004, GarciaMunoz_2007,Murray-Clay_2009,Owen&Jackson_2012, Owen_2019}. The resulting low-density, high-altitude layers are best probed at ultraviolet (UV) wavelengths, which provide access to strong resonance transitions of atomic and ionized species that can trace both composition and mass loss.

UV transmission spectroscopy with the Hubble Space Telescope (HST) has enabled detections of metals and ions in several close-in giant planets, including species such as C\,{\sc ii}, Fe\,{\sc ii}, O\,{\sc i}, Mg\,{\sc i}/Mg\,{\sc ii}, and Si\,{\sc iii}. Many of these measurements were obtained with the Space Telescope Imaging Spectrograph (STIS) \citep[e.g.,][]{Vidal-Madjar_2003,Vidal-Madjar_2004,Vidal-Madjar_2013,Sing_2019,Cubillos_2020,Ben-Jaffel_2022,Baldwin_2026} and the Cosmic Origins Spectrograph (COS) \citep[e.g.,][]{Fossati_2010,Linsky_2010,Ben-Jaffel&BAllester_2013,Ballester&Ben-Jaffel_2015,GarciaMunoz_2021,Ben-Jaffel_2022,DosSantos_2023}, demonstrating that UV lines can reveal extended atmospheres and provide constraints on the physical state of the escaping gas.

Ultra-hot Jupiters (UHJs) are particularly favorable targets for these studies because their extreme temperatures fundamentally alter atmospheric chemistry and opacity. At temperatures above 2,000 K, molecules thermally dissociate, atoms become efficiently ionized, and condensation of refractory elements is suppressed, leaving abundant gaseous metals (e.g., Fe and Mg) available as tracers of photoionization, dynamics and escape \citep{Visscher_2010,Wakeford_2017,Linssen&Oklopcic,Lothringer_2020}. Observational surveys and theoretical work have consequently revealed a diverse inventory of atomic species in UHJ atmospheres \citep{Lothringer_2018,Parmentier_2018}, and predict that these planets may exhibit strong signatures of upper-atmosphere heating and mass loss.

In this study, we present new observations of WASP-76b \citep{West_2016}, complementing the existing HST transmission spectrum presented by \cite{Fu_2021}. WASP-76b is an UHJ orbiting an F-type star ($V_\mathrm{mag}$ = 9.52, $T_\mathrm{eff}$ $\sim$ 6366 K) every 1.81 days at a separation of 0.033 AU (see \autoref{tab: sys_params} for all the system parameters). It is a benchmark planet that has been extensively studied from the ground and from space, with numerous atomic, ionized and molecular species reported in its atmosphere: Ba\,{\sc ii}, Ca\,{\sc i}, Ca\,{\sc ii}, Co\,{\sc i}, Cr\,{\sc i}, Fe\,{\sc i}, Fe\,{\sc ii}, H\,{\sc i}, He\,{\sc i}, K\,{\sc i}, Li\,{\sc i}, Mg\,{\sc i}, Mn\,{\sc i}, Na\,{\sc i}, Ni\,{\sc i}, O\,{\sc i}, Sr\,{\sc ii}, V\,{\sc i}, CO, H$_2$O, HCN, OH, TiO and VO (see \citealt{CostaSilva_2024}, for more details on these detections). 

We motivate this work with prior observations from ESPRESSO \citep{Ehrenreich_2020} and HST \citep{Fu_2021} which reveal high-altitude atomic metals and complex thermal structure on WASP-76b. \cite{Ehrenreich_2020} report an asymmetrical detection of neutral iron (Fe\,{\sc i}) in the optical transmission spectrum that is present on the evening (dayside) terminator but is absent on the morning terminator; this absorption is also blueshifted and stronger during egress than during ingress. This pattern in velocity and phase-resolved strength was interpreted as nightside condensation (or ``iron rain") removing gaseous Fe on the morning limb, combined with zonal winds to produce the observed Doppler shifts. \cite{Fu_2021} extended the picture to 0.29 $-$ 4.5 $\mu$m, finding a steep excess in transit depth at NUV/blue wavelengths that cannot be explained by Rayleigh scattering or hazes alone and requiring neutral and ionized metals (e.g. Fe, Ti, V, Ni, SiO, Ca) to fit the data. They further show that heavy metals, H$_2$O and TiO seen in transmission together with muted H$_2$O emission but strong CO emission at 4.5 $\mu$m, are consistent with an inverted dayside T-P profile. \cite{Fu_2021} advocates for the inclusion of NUV observations to identify high-altitude heavy metal species, link composition to upper-atmosphere heating and thermal inversion and test condensation/rainout and ionization processes at the terminator. Subsequent ground-based high-resolution studies have reported additional metal detections in WASP‑76b \citep[e.g.,][]{Tabernero_2021,Kesseli_2022,AzevedoSilva_2022}, further motivating targeted NUV follow-up.  Motivated by these results, we aimed to determine whether metals are present in WASP-76b's NUV spectrum and if they are escaping.

In this paper, we present a NUV transmission spectrum of WASP-76b from HST (GO-16724, PI: L. Dos Santos). Our scientific objectives are to search for signatures of atmospheric escape and assess whether any outflow is consistent with hydrodynamic escape, evaluate hydrodynamic models against the observed transmission spectrum, estimate the atmospheric mass-loss rate, and determine whether metals extend to the Roche lobe. The paper is organized as follows: \autoref{sec:Observations/Analysis} describes the observations and data analysis; the results are shown in \autoref{sec:Results}; \autoref{sec:Modeling} shows the modeling of the lower and upper atmosphere. A discussion and implications is provided in \autoref{sec:Discussion} and conclusions in \autoref{sec:Conclusion}. 

\begin{table}[h]
\caption{Stellar and Planetary Parameters of the WASP-76 system.}
\vspace{-0.1cm}
\hspace{-1.6cm}
\centering
\begin{tabular}{c c}
\hline\hline
Stellar Parameters & Value\\
\hline
Radius ($R_s$) & 1.744$_{-0.042}^{+0.045}$ $R_\sun$\\
Mass ($M_\star$) & 1.467 $_{-0.081}^{+0.079}$  $M_\sun$\\ 
Density ($\rho_*$) & 0.391 $_{-0.024}^{+0.018}$  g/cm$^3$\\
Eff. temperature ($T_\mathrm{eff}$) & 6366 $\pm$ 90  K \\
Surface gravity ($\log g_*$) & 4.123 $_{-0.020}^{+0.016}$ cm/s$^2$\\
Metallicity [Fe/H] & 0.204 $_{-0.096}^{+0.093}$ dex\\
\hline
Planetary Parameters & Value\\
\hline
Radius (\(R_p\)) & 1.845$_{-0.046}^{+0.050}$ $R_J$\\
Mass (\( M_p \)) & 0.910 $\pm$ 0.042 $M_J$\\
Orbital period ($P$) & 1.80988158 $\pm$ 0.00000030 days\\
Scaled semi-major axis ($a/R_s$) & 4.078$_{-0.083}^{+0.060}$\\
Transit center time ($T_0$) & 2457360.29300 $\pm$ 0.00014 BJD\\
Impact parameter ($b$) & 0.103$_{-0.071}^{+0.084}$\\
Eccentricity ($e$) & 0.016$_{-0.011}^{+0.018}$\\ 
Inclination ($i$) & 88.5$_{-1.2}^{+1.0}$ deg\\
Argument of periastron ($\omega$) & 62.0$_{-82.0}^{+67.0}$ deg\\
Eq. Temperature ($T_\mathrm{eq}$) & 2231$_{-36}^{+37}$ K\\
Semi-major axis ($a$) & 0.03304$_{-0.00062}^{+0.00058}$ AU\\
\hline
\end{tabular}
\small Reference: \citet{Fu_2021} \\
\label{tab: sys_params}
\end{table}

\section{Observations and Data Analysis} \label{sec:Observations/Analysis}

\subsection{Observations} \label{sec:Observations}

\begin{figure*}
    \includegraphics[width=1 \textwidth]{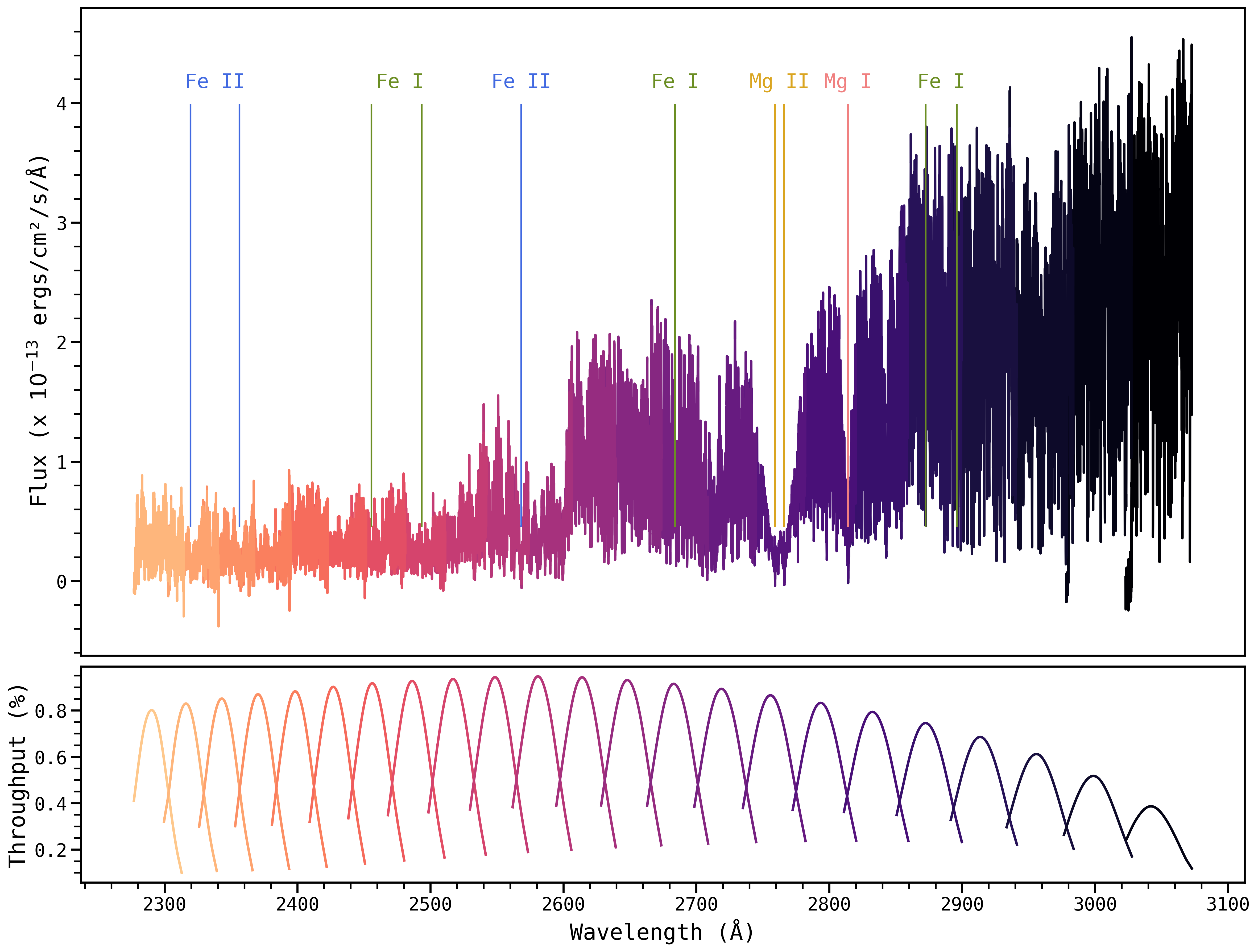}
    \caption{\textit{Top:} Flux calibrated out-of-transit spectrum of WASP-76b with each order plotted in a different color. Resonant metal lines of Mg and Fe are indicated. \textit{Bottom:} Instrumental throughput of each order.}
    \label{fig:1}
\end{figure*}

We successfully observed two transits of WASP-76b using the Hubble Space Telescope Imaging Spectrograph (HST STIS) instrument with the NUV-Multi-Anode Microchannel Array (NUV-MAMA) detector using the E230M echelle grating. This grating has a 0.2"$\times$0.2" aperture and a resolving power of R = 30,000. The configuration setting was centered at 2707 {\AA}, providing a wavelength coverage from 2274 $-$ 3119 {\AA} across 24 orders, however, we discarded the last order following \citet{Sing_2019} (see \autoref{fig:1}). The resulting spectra exhibited a dispersion of approximately 0.049 {\AA} per pixel on the detector. Each transit was observed in a single visit on January 12, 2022 (visit A) and again on January 14, 2022 (visit B), where the latter attempt failed. This observation was subsequently repeated on December 28, 2022 (visit C), which also resulted in failure. A successful observation of the second visit was achieved on February 8, 2023 (visit D). These outcomes were the result of guide star failure, the observations log can be seen in \autoref{tab: obs_log}. For both transits, our aim was to cover five consecutive HST orbits with the transit event occurring in the third and fourth orbits; however, for visit D we only achieved four orbits. Due to strong systematics and their effect on the continuum, we only used visit D for the spectrophotometric analyses i.e., light-curve extraction in wavelength bins, while we incorporated visit A and D for the single line analysis since measurements are normalized to the continuum (see details in \autoref{Appendix: Raw Visits}). The spectra were obtained in TIME-TAG mode, where each photon's position and detection time are recorded in an event list, with detection times captured with a precision of 125 $\mu$s.

\begin{table*}
\caption{Observations log of WASP-76b transits with HST/STIS.}
\centering
\begin{tabular}{c c c c c}
\hline\hline
Visit & Dataset & Start Time (UT) & Exp. Time (s) & Phase\\
\hline
A & oep501010 & 2022-01-12 05:39:26.754 & 1563.20 & out-of-transit\\
 & oep501020 & 2022-01-12 07:10:16.755 & 1792.20 & ingress\\
 & oep501030 & 2022-01-12 08:45:34.754 & 1792.20 & in-transit\\
 & oep501040 & 2022-01-12 10:20:53.755 & 1792.20 & near egress\\
 & oep501050 & 2022-01-12 11:56:12.755 & 1792.20 & out-of-transit\\
\hline
D & oep5s2010 & 2023-02-08 23:30:37.806 & 2079.20 & out-of-transit\\
 & oep5s2020 & 2023-02-09 00:58:00.804 & 2496.20 & ingress\\
 & oep5s2030 & 2023-02-09 02:33:09.804 & 2496.20 & in-transit\\
 & oep5s2040 & 2023-02-09 04:08:20.805 & 2496.20 & near egress\\
\hline
\end{tabular}
\label{tab: obs_log}
\end{table*}

\subsection{Broadband Data Analysis} \label{sec:Broadband analysis}

\subsubsection{Data reduction} \label{sec:Data Reduction}
To start the reduction, we divided the spectrum of each visit into seven sub-exposures using \texttt{stistools} 1.4.5 which yielded a total of 28 sub-exposures for visit D and 35 sub-exposures for visit A. Afterwards, we used \texttt{calstis} version 4.3.2 to perform calibration steps, then applying flat field corrections, and then perform box spectral extraction and finally calibrating wavelengths.

We used the \texttt{stissplice}\footnote{\footnotesize{Available in \url{https://github.com/ladsantos/stissplice}.}} software library to handle overlapping spectral orders. It first reads the 1D spectrum data from FITS files and then identifies unique and overlapping sections of the spectrum. The library then merges these overlapping sections using a weighted averaging technique that accounts for signal-to-noise ratios (SNR) and data quality flags. Lastly, the unique and merged spectral components are concatenated to produce a comprehensive spliced spectrum. 

\subsubsection{Systematics correction} \label{sec:Systematics}

Spectrophotometric light curves obtained with the STIS instrument on HST are significantly influenced by systematic effects related to the instrument. To handle systematics and light curve fitting, we follow certain portions of the methodology described in \cite{Allen_2024}, which implements the pipeline from the Sculpting Hubble’s Exoplanet Legacy (SHEL) program (HST GO-16634, PI: D. Sing) \footnote{\footnotesize{Publicly available at \url{https://github.com/natalieallen/stis_pipeline}.}}. A summary of the approach is briefly given below. We start with the spliced ({\tt *\_x1d.fits}) files as well as the engineering jitter files ({\tt *\_jit.fits}) to extract the respective time-series exposures. 

Similarly to \cite{Sing_2019}, we included the measured parameters that describe HST's Pointing Control System during an observation period in our systematic effects correction model. These parameters are generated by the Engineering Data Processing System (EDPS). Because the jitter telemetry is sampled more finely than our science sub-exposures, we time-bin (mean) each jitter vector over the time span of each sub-exposure to match the light-curve cadence. From all the parameters available we selected Latitude and Longitude plus the six jitter parameters that exhibited the highest correlation with the white light curve for detrending: \texttt{V2\_dom, V2\_roll, V3\_roll, SI\_V2\_avg, RA, DEC, Latitude} and \texttt{Longitude} (see \autoref{tab: jitter_corr}). A description of the jitter parameters used is shown in \autoref{tab: jitter_all}. To decorrelate these detrending vectors, we initially normalize each measured quantity by subtracting the mean and subsequently dividing by the standard deviation.

\begin{table}[h]
    \caption{Correlation between jitter data and white light curve fluxes.}
    \centering
    \begin{tabular}{c c c}
    \hline\hline
    Parameters & Spearman-$r$ & $p$-value\\
    \hline
    \texttt{V2\_dom} & 0.49186 & 0.00785\\
     \texttt{V2\_roll} & -0.49316 & 0.00766\\ 
    \texttt{V3\_roll} & -0.48878 & 0.00831\\
    \texttt{SI\_V2\_avg} & -0.41325 & 0.02883\\
    \texttt{RA} & -0.46689 & 0.01226\\
    \texttt{DEC} & 0.47345 & 0.01093\\
     \texttt{Latitude} & -0.09250 & 0.63967\\ 
    \texttt{Longitude} & 0.11658 & 0.55465\\ 
    \hline
     \texttt{V3\_dom} & -0.15107 & 0.44287\\
    \texttt{SI\_V2\_rms} & -0.25068 & 0.19821\\
    \texttt{SI\_V2\_p2p} & -0.29283 & 0.13047\\
    \texttt{SI\_V3\_avg} & 0.12151 & 0.53793\\
    \texttt{SI\_V3\_rms} & 0.10126 & 0.60816\\
    \texttt{SI\_V3\_p2p} & -0.13848 & 0.48221\\
    \texttt{Roll} & 0.00712 & 0.97133\\
    \texttt{Limbang} & 0.03612 & 0.85519\\
    \texttt{Termang} & 0.21272 & 0.27713\\
    \texttt{LOS\_zenith} & -0.03612 & 0.85519\\
    \texttt{Mag\_V1} & 0.14669 & 0.45636\\
    \texttt{Mag\_V2} & -0.10783 & 0.58496\\
    \texttt{Mag\_V3} & -0.10947 & 0.57923\\
    \hline
    \end{tabular}
    \label{tab: jitter_corr}
\end{table}

We detrend systematics and fit the white and spectral light curves using \texttt{juliet} \citep{juliet} and the same set of jitter regressors, adopting two alternative systematics treatments: (i) a Gaussian process (GP) model and (ii) a linear model. In both cases we include standard \texttt{juliet} nuisance parameters \textit{mflux} (the offset relative flux for the photometric instrument) and \textit{sigma$\_$w} (a jitter term added in quadrature to the instrument uncertainties). For the GP detrending, we use a Matern-3/2 kernel as implemented in \texttt{george} \citep{george},
\begin{equation}
k(\mathbf{x},\mathbf{x}')=\sigma_{\rm GP}^{2}\left(1+\sqrt{3}\,r\right)\exp\!\left(-\sqrt{3}\,r\right),
\end{equation}
with
\begin{equation}
r=\sqrt{\sum_{i=1}^{d}\alpha_i\,(x_i-x_i')^{2}},
\end{equation}
where $\mathbf{x}$ is the vector of standardized detrending regressors, $\sigma_{\rm GP}$ corresponds to \textit{GP\_sigma}, $\alpha_i$ corresponds to \textit{GP\_malpha}$_i$ and $d$ is the number of regressors (here $d=8$). Equivalently, $\alpha_i = 1/\ell_i^{2}$, where $\ell_i$ is the GP correlation length scale associated with regressor $i$. We use non-informative priors adopted in \citet{Allen_2024}, $\textit{mflux}\sim\mathcal{N}(0,1)$, $\textit{sigma\_w}\sim\log\mathcal{U}(0.1,10^4)$ for the GP fits ($\textit{sigma\_w}\sim\log\mathcal{U}(0.1,10^3)$ for the linear fits), and $\textit{GP\_sigma}\sim\log\mathcal{U}(10^{-6},10^{6})$. We place an exponential prior on each $\alpha_i$, $\alpha_i\sim\mathrm{Exp}(\lambda=1)$, where an ``exponential prior of 1” denotes an exponential distribution with rate $\lambda=1$ (mean 1). For the linear treatment, each regressor coefficient $\theta_i$ (implemented as \textit{theta}$_i$) is given a broad uniform prior $\sim\mathcal{U}(-100,100)$.

\begin{figure*} [ht]
    \centering
    \includegraphics[width=1\textwidth]{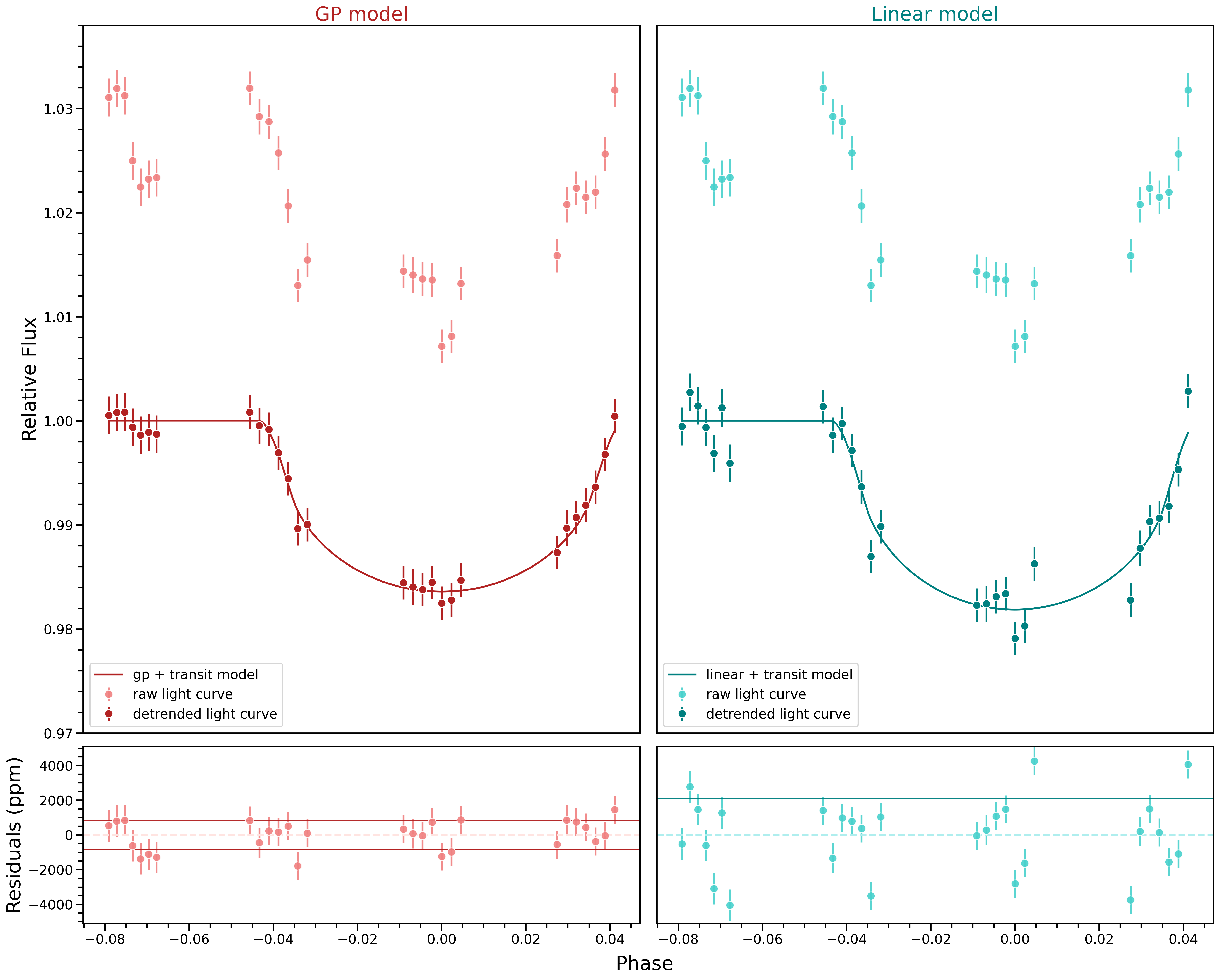}
    \caption{WASP-76b white light curves for visit D with GP model (left) and linear model (right) with offset. \textit{Top:} The raw and detrended white light curves (burgundy and teal). The best-fit light curve model is overplotted. \textit{Bottom:} Corresponding residuals from the white light curve fit are shown in the colored circles. The solid lines correspond to one times the standard deviation.}
    \label{fig:2}
\end{figure*}

\begin{figure*} [ht]
    \centering
    \includegraphics[width=1\textwidth]{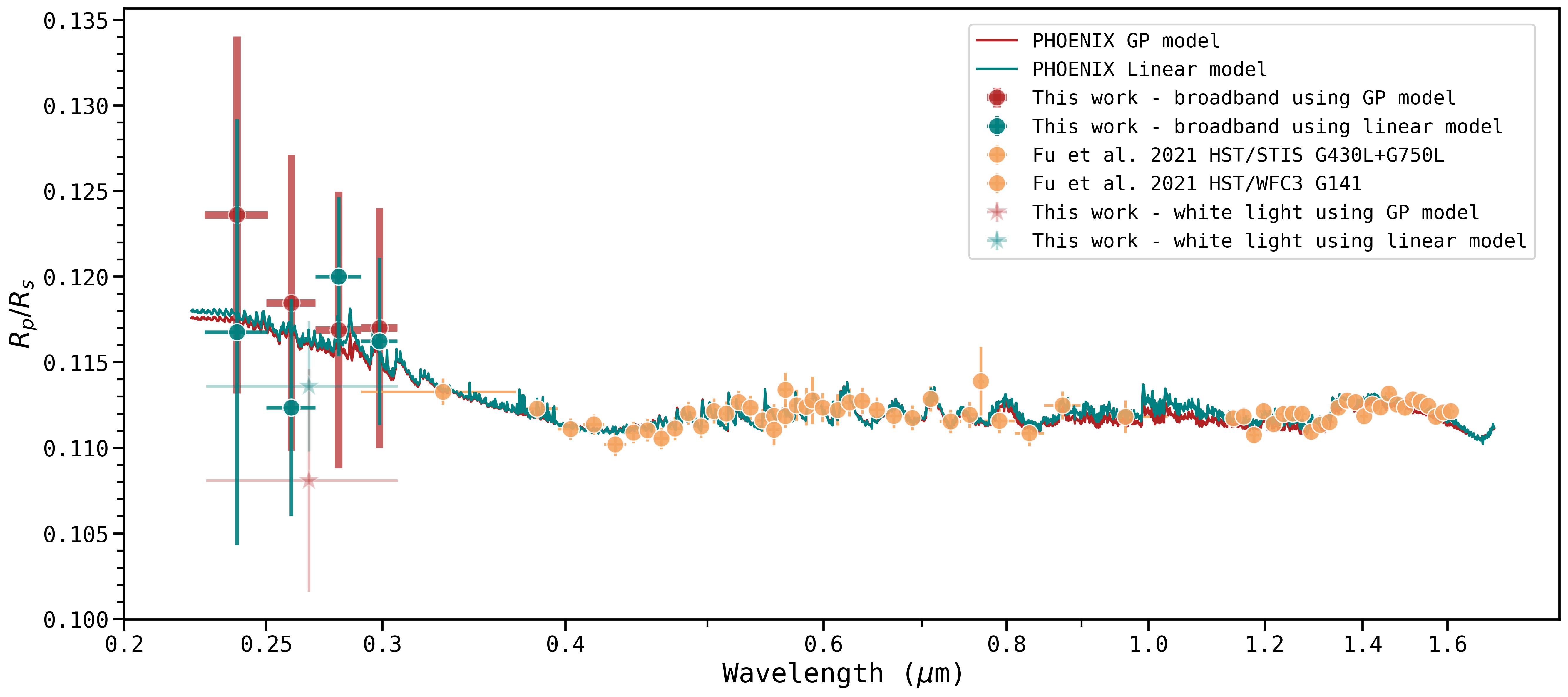}
    \caption{The WASP-76b HST transmission spectrum. The NUV values come from the analysis of visit D using HST/STIS E230M (burgundy and teal) using $\sim$ 200 {\AA} wide bins. The optical and NIR observations are from the analysis of \cite{Fu_2021} of HST/STIS G430L $+$ G750L and HST/WFC3 G141 (sand). Star symbols represent the white light curve values and the best-fit \texttt{petitRADTRANS} models are shown in solid lines for each systematics treatment.}
    \label{fig:3}
\end{figure*}

\subsubsection{White Light Curve Fitting} \label{sec:White light}

To derive a white light curve, we integrate the flux over the entire wavelength range, specifically from 2274.5 to 3072.5 {\AA} for each exposure. The exposure time stamps provided in JD$_{\mathrm{UTC}}$ are converted to BJD$_{\mathrm{TDB}}$ using the \texttt{barycorrpy} framework \citep{Kanodia_2018}. Each white light curve is normalized relative to the median flux of the out-of-transit data collected during each visit, and for fitting the light curves, we employ  \texttt{juliet}. The modeled flux measurement incorporates both a geometric transit model that accounts for limb darkening and a systematic error correction model, the latter of which is detailed in \autoref{sec:Systematics}. This study utilizes stellar and planetary parameters from \cite{Fu_2021} (see \autoref{tab: sys_params}). During the transit fitting process, we fixed all system parameters except for $R_p/R_s$, while fitting all coefficients of the systematic detrending vectors. We also fit for \textit{mflux} and \textit{sigma$\_$w} for both detrending methods and additionally for \textit{GP$\_$sigma} in the GP case. For limb darkening, we adopted the quadratic law, calculating the coefficients as detailed in \cite{Sing_2010}. For the model, we employed a 3D stellar model from the Stagger-grid similar to \cite{Sing_2019}, specifically with parameters $T_\mathrm{eff}$ = 6366 $\pm$ 90 K, [Fe/H] = 0.204$_{-0.096}^{+0.093}$ dex and $\log g_*$= 4.123$_{-0.020}^{+0.016}$ cm/s$^2$ corresponding to the WASP-76 system. The coefficients were subsequently computed across the specified wavelength range, encompassing the entire spectrum for the white light curve and the particular binning ranges selected for the spectroscopic fits (see next section). Following the calculation of these coefficients, we applied the quadratic limb darkening law parameterization as outlined by \citet{Kipping_2013} for parameters q$_1$ and q$_2$, which were fixed for the rest of the study. We sample posterior distributions with \texttt{dynamic dynesty} to obtain robust parameter estimates and uncertainties. The resulting white light fits can be seen in \autoref{fig:2}.

\subsubsection{Spectral Light Curve Fitting} \label{sec:spectroscopic fits}
To generate spectroscopic light curves for the E230M observations, we segment the data into four spectrophotometric channels covering the range of 2273.5 to 3072.5 {\AA}. The binning strategy was devised to optimize the SNR (SNR \textgreater 100 per bin) while maintaining appropriate spectral resolution; it follows the binning scheme of \cite{Sing_2019}. Leveraging the parameters from the white light curve fitting, we implement an analogous fitting procedure for the spectroscopic light curves. In this analysis, we impose constraints on all system parameters except for $R_p/R_s$, while simultaneously fitting the coefficients of the detrending vectors. Notably, we utilize the coefficients derived from the white light curve fitting as informative priors for the spectroscopic fits. For the limb darkening coefficients, we calculate the values pertinent to each bin and subsequently fix these parameters during the analysis. For the GP and linear systematics models, we use the same procedure as it was described in the white light curve fitting section. \autoref{fig:3} shows our resulting broadband transmission spectrum.

\subsubsection{Analysis overview}
We pursue two complementary analyses. First, a spectrophotometric approach, in which we fit wavelength-binned light curves to infer $R_p/R_s$ (broadband analysis,  $\sim$ 200 {\AA} bins, and a narrow-band analysis, of 4 {\AA} bins). Second, a velocity-resolved single line analysis, in which we compare in-transit and out-of-transit spectra in the stellar rest frame to localize absorption within specific line profiles. The single line analysis results are then used to define targeted sub-{\AA} spectrophotometric bins (e.g., Mg\,{\sc ii} in the blue wing), enabling direct comparison to the broadband optical radius and the Roche-lobe scale (see more details on these procedures in \autoref{sec:Results}).

\section{Results} \label{sec:Results}

\subsection{White Light Curve Analysis}

We applied our detrending method described in \autoref{sec:Systematics} to the data gathered during the transit of WASP-76b. Visit A is highly affected by systematic effects (see left panel of \autoref{fig:10} in \autoref{Appendix: Raw Visits}) and we do not include it on this part of the analysis whereas visit D has better quality, which is less affected by systematics and is the one analyzed in this section. The raw, systematic-corrected light curves and best fitting models using the GP and linear models are shown in \autoref{fig:2}. The systematics were fit using the eight detrending vectors listed in \autoref{tab: jitter_corr}, the derived coefficients of the jitter parameters can be found in \autoref{tab: jitter_coeff}. The resulting $R_p/R_s$ for each model can be found in \autoref{tab: broadband_spec}; we see that the values agree with each other within their respective uncertainties. Our new measurements of $R_p/R_s$ are consistent with the previously published value for WASP-76b ($R_p/R_s$ = 0.1087$\pm$0.0005) from \cite{Fu_2021}. 

\subsection{Broadband Analysis} \label{sec: Results- broadband}

For the broadband analysis, we used the same bin width configuration of $\sim$ 200 {\AA} as in \cite{Sing_2019}. We fitted the broadband data using the same detrending models as the one used to correct the white light curve. These results are shown in \autoref{fig:3} and \autoref{tab: broadband_spec}.  We see an increase in $R_p/R_s$ towards the shorter wavelengths for the linear model, however, with the relatively large uncertainties for each model, it is difficult to decisively conclude that there are absorbers at those wavelengths. Since upper atmospheric signals can be narrow, the next step is to conduct a high-resolution analysis focused on the lines of interest (see \autoref{sec: Results - high res}).

\begin{table*}
\caption{WASP-76b broadband transmission spectrum results with quadratic limb darkening coefficients.}
\centering
\begin{tabular}{c c c c c c c}
\hline\hline
Bands & $\lambda_c$ (\AA) & $\Delta\lambda$ (\AA) &
\multicolumn{2}{c}{$R_p/R_s$} & \multicolumn{2}{c}{Limb darkening coefficients} \\
 &  &  & GP & Linear & a & b \\
\hline
White light & 2673 & 799 & $0.1081\pm0.0065$ & $0.1136\pm0.0038$ & 0.8496 & 0.0444 \\
Bin 1       & 2387 & 236 & $0.1137\pm0.0113$ & $0.1136\pm0.0128$ & 1.2023 & -0.2843 \\
Bin 2       & 2600 & 200 & $0.1081\pm0.0095$ & $0.1091\pm0.0065$ & 0.9145 & -0.0027 \\
Bin 3       & 2800 & 200 & $0.1064\pm0.0089$ & $0.1170\pm0.0048$ & 0.8287 & 0.0524 \\
Bin 4       & 2986 & 172 & $0.1065\pm0.0077$ & $0.1131\pm0.0050$ & 0.6453 & 0.2340 \\
\hline
\end{tabular}
\label{tab: broadband_spec}
\end{table*}

\subsection{Narrow-band Analysis} \label{sec: Results - narrowband}

\begin{figure}[h]
    \includegraphics[width=0.49 \textwidth]{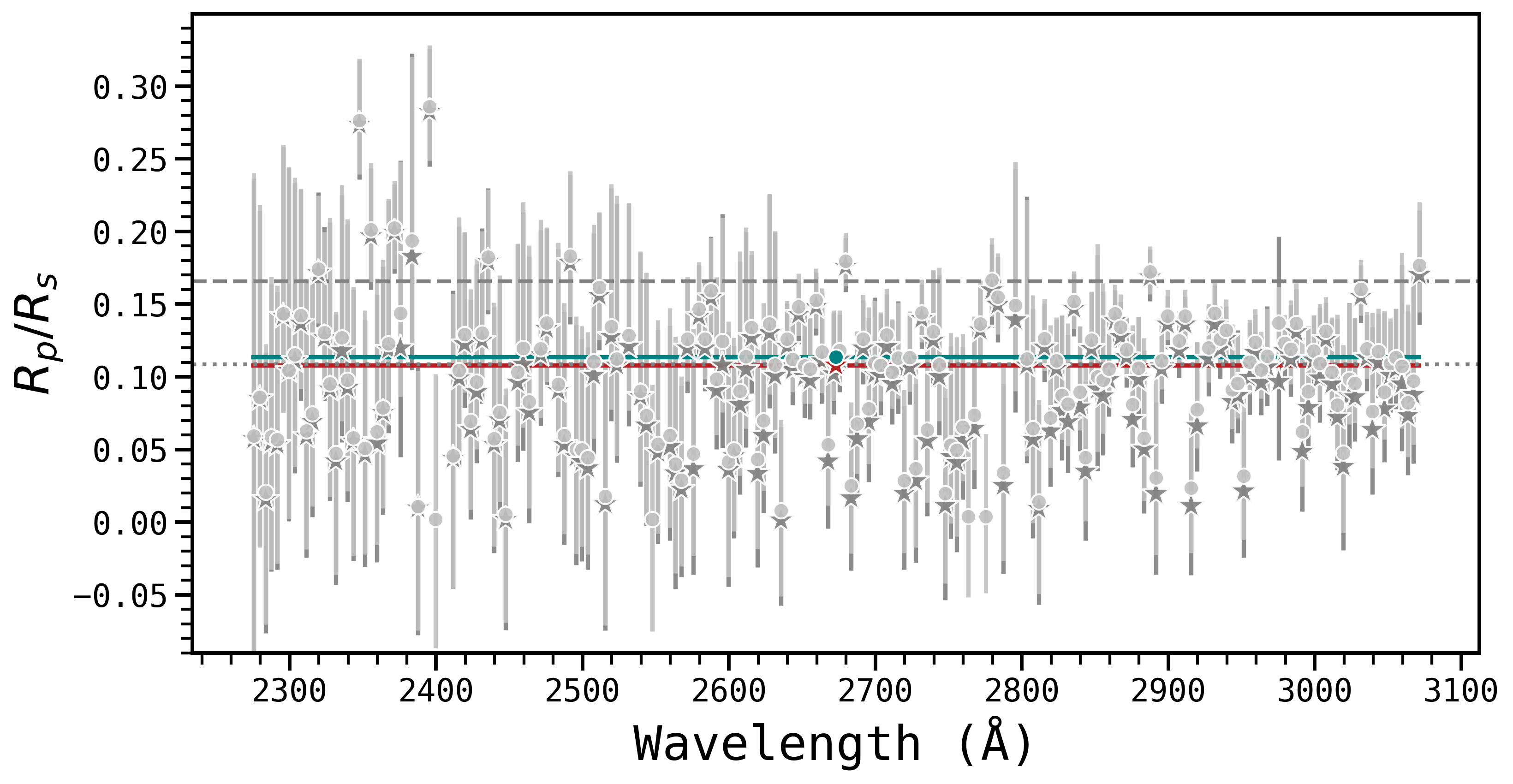}
    \caption{NUV transmission spectrum of WASP-76b for visit D using 4 {\AA} wide bins. The dark gray points represent the spectrum derived from the GP model, while the lighter gray x's correspond to the linear model. The transmission spectrum for the white light curve is indicated in burgundy as $R_p/R_s$ (NUV) = 0.1081$\pm$0.0065 and in teal as $R_p/R_s$ (NUV) = 0.1136$\pm0.0038$ for each respective model. The optical radius from \cite{Fu_2021} is shown in the dotted line and the the dashed line indicates the Roche-lobe scale.}
    \label{fig:4}
\end{figure}

To explore the presence of metallic species in the upper atmosphere, we determined the transmission spectrum in narrow bands of 4 {\AA} bins (see \autoref{fig:4}) as in \cite{Sing_2019}. The first indication would be an increase in the apparent radius of the planet at wavelengths that would correspond to heavy metal species at these high altitudes or if they are escaping the atmosphere. We see in \autoref{fig:4}, values of $R_p/R_s$ that extend beyond the Roche-lobe radius $R_{\rm RL}/R_s$ = 0.1659$_{-0.0042}^{+0.0041}$  (see \autoref{sec:Discussion}), however, none of them correspond to known metallic species that absorb in those wavelengths. Given the large uncertainties and lack of correspondence with known strong transitions, we interpret these excursions as statistical scatter. We see a high absorption at 2383.5 {\AA}, which could correspond to a Fe\,{\sc ii} line typically found at 2382 {\AA}. Nevertheless, when we look at these lines in high resolution, there is no detection of excess absorption of Fe\,{\sc ii} (see \autoref{sec: Results - high res}). We note that the resulting narrow-band transmission spectrum does not provide any evidence of absorption on the species of interest: Mg\,{\sc i}, Mg\,{\sc ii}, Fe\,{\sc i}, and Fe\,{\sc ii}. Additionally, we can see that there is higher variability in $R_p/R_s$ in the shorter wavelengths but there are no apparent trends visible across the spectrum. 

\subsection{Single Line Analysis - Mg\,{\sc i}, Mg\,{\sc ii}, Fe\,{\sc i} and Fe\,{\sc ii}} \label{sec: Results - high res}

\begin{figure*}[ht]
    \includegraphics[width=1 \textwidth]{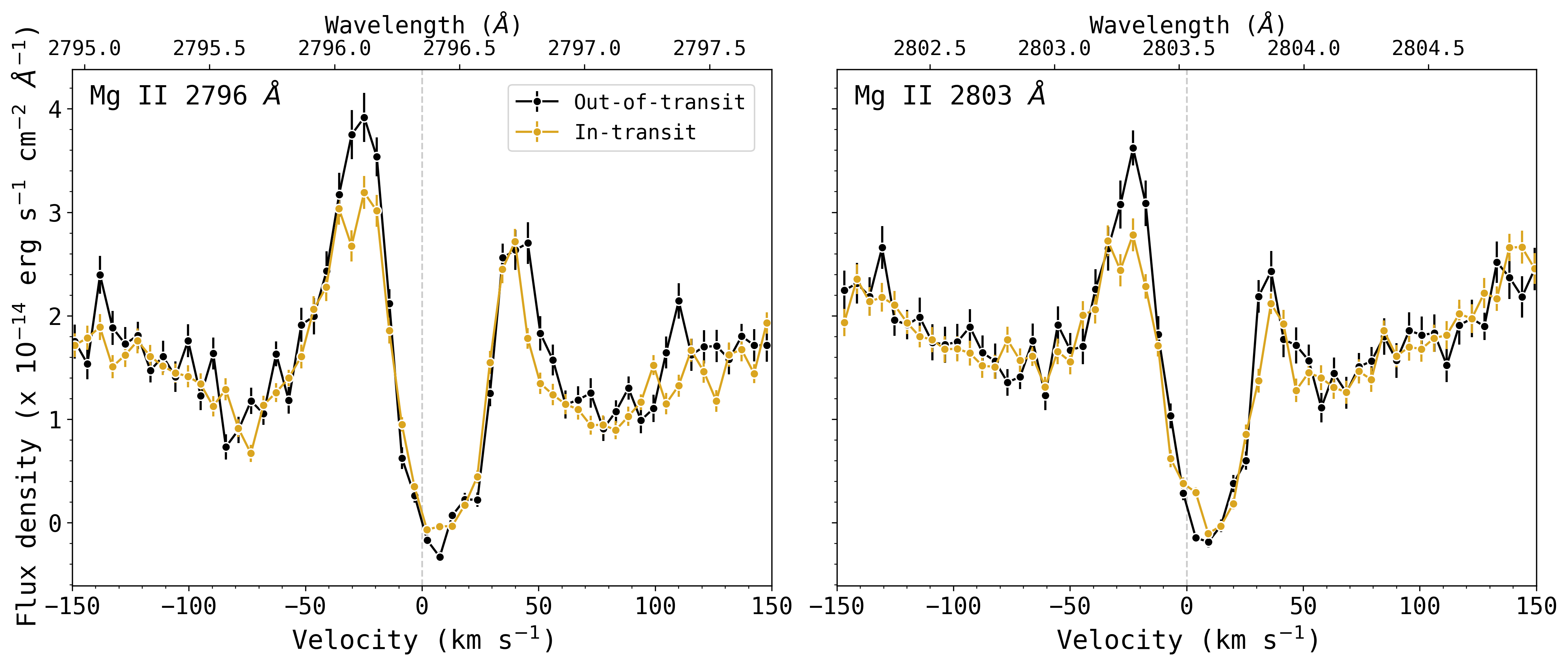}
    \caption{Mg\,{\sc ii} resonance doublet in the stellar rest frame. Left: $k$ line at 2796.35 {\AA}; right: $h$ line at 2803.53 {\AA}. The out-of-transit points are shown in black while golden points show the in-transit spectrum (visit A and D). The dashed vertical line marks the line center ($v = 0$). Both lines show excess absorption in the blue wing; we quantify the signal by integrating over [-50, -10] km s$^{-1}$.}
    \label{fig:5}
\end{figure*}

Due to our broadband and narrow-band searches not yielding significant detections of metal species absorptions -- most likely because our expected signals are below our detection threshold given the low SNR -- we carried out a single line analysis using visit A and visit D. In this approach, we compare in-transit and out-of-transit spectra specifically in the lines of interest, after transforming to the stellar rest-frame velocity scale as well as wavelength space (see \autoref{fig:5}, \autoref{fig:11}, \autoref{fig:12}, \autoref{fig:13}).

We start by exploring the Mg lines. Mg\,{\sc i} has a central wavelength of 2852.965 {\AA} and the Mg\,{\sc ii} doublet has central wavelengths at 2796.35 {\AA} and 2803.53 {\AA}. The Mg\,{\sc ii} 2796.35 {\AA} transition is commonly denoted the $k$ line, and the 2803.53 {\AA} transition the $h$ line; we adopt this notation throughout. We find no evidence for absorption by Mg\,{\sc i} (see \autoref{fig:11}). In contrast, we detect and resolve both components of the Mg\,{\sc ii} doublet. This is consistent with expectations: the first ionization potential (IP) of Mg is only $\sim$ 7.6 eV, meaning Mg is efficiently photoionized in the upper atmosphere of an UHJ, analogous to the behavior of carbon and other low-IP species \citep[e.g.,][]{ GarciaMunoz_2007}. Mg\,{\sc ii} is therefore the likely dominant ionization state at the altitudes probed by NUV resonance lines (see discussion in \autoref{sec: Discussion - ionization}). The in-transit spectra show excess absorption relative to out-of-transit spectra in the blue wing of both lines, over Doppler velocities [-50, -10] km s$^{-1}$ (see \autoref{fig:5}; quantitative measurements are given below). The stellar flux approaches zero near the line center, so our data provide limited information at low velocities.

\begin{figure*}[ht]
    \includegraphics[width=1 \textwidth]{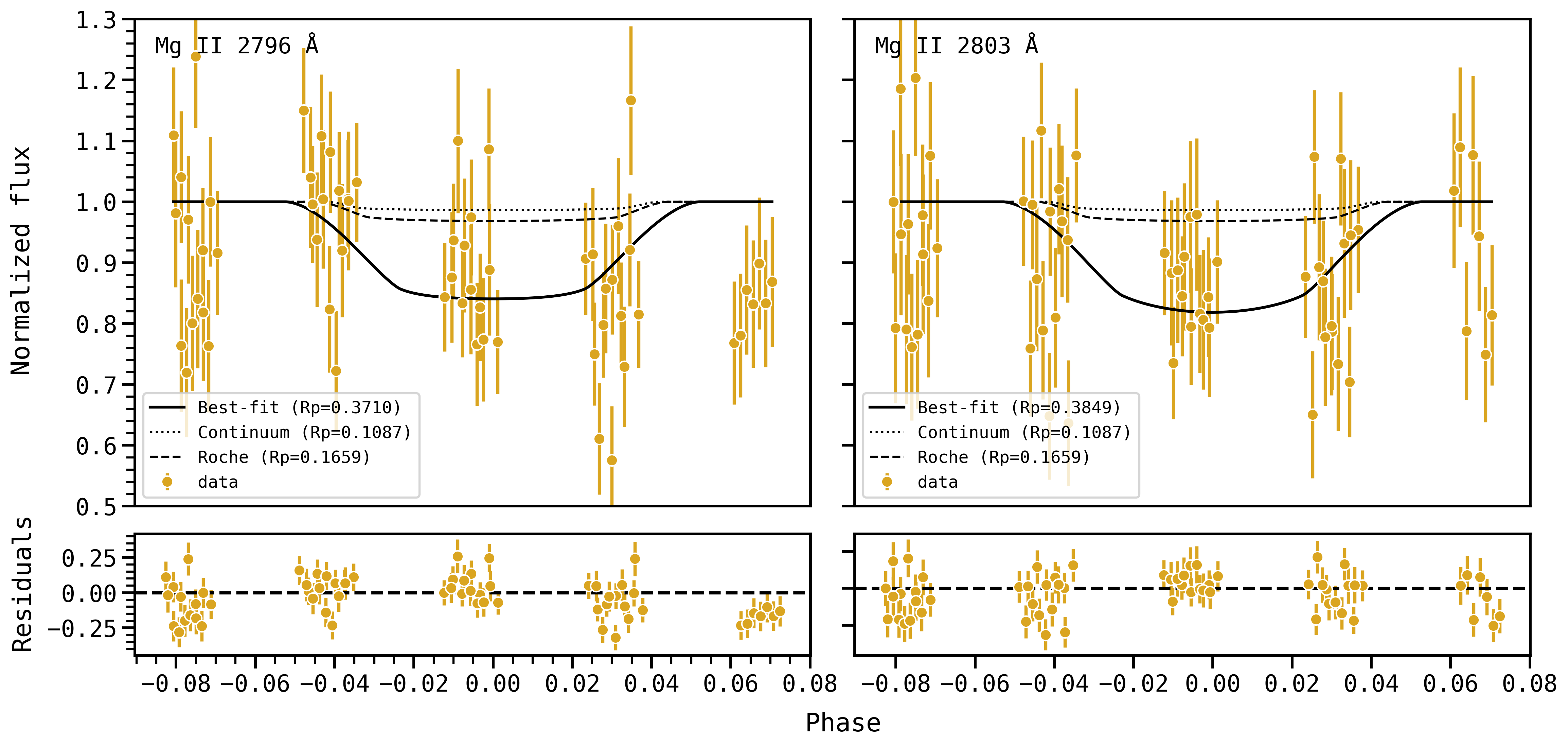}
    \caption{Mg\,{\sc ii} light curves constructed by integrating the flux in the blue wing velocity interval [-50, -10] km s$^{-1}$ for the $k$ line (2796.35 {\AA}; left) and the $h$ line (2803.53 {\AA}; right). Gold points show the normalized, co-added time series from visit A and D. The solid black curve is the best-fit gray transit radius model, parameterized by an effective radius ratio $R_{\rm eff}/R_s$. For comparison, the dotted curve shows the expected transit radius in the optical from \citet{Fu_2021}, and the dashed curve shows a ``Roche-lobe transit'' corresponding to an occulting disk with radius $R_{\rm RL}/R_s$. Bottom panels show the residuals relative to the best-fit model.}  
    \label{fig:6}
\end{figure*}

We construct Mg\,{\sc ii} light curves using the \texttt{RWDDT Utils} package\footnote{\footnotesize{Publicly available at \url{https://github.com/spacetelescope/rocky-worlds-utils}.}}, which is capable of integrating fluxes in fractional pixels and properly calculating uncertainties of STIS UV data using Poisson statistics. This is necessary when integrating fluxes at only a few pixels since the uncertainties are dominated by Poisson counting noise. The resulting light curves are shown in \autoref{fig:6}. Because these light curves trace resonance-line absorption in a line-dominated bandpass (and they are restricted in velocity space), we parameterize the signal using an effective radius ratio, $R_{\rm eff}/R_s$, defined as the radius ratio of a gray transit model that reproduces the observed bandpass light-curve depth. We fit each light curve with \texttt{batman} \citep{batman}, fixing all the system parameters (\autoref{tab: sys_params}) and adopting the same limb darkening procedure described in previous sections while allowing $R_{\rm eff}/R_s$ to vary. 

To avoid bias from any possible post-transit absorption, we apply a visit-specific normalization using pre-transit data before co-adding the light curves. For visit D, we normalize against the first HST orbit, which is fully pre-transit. For visit A, due to strong effects of slit losses, we utilize a different strategy to measure the Mg\,{\sc ii} excess absorption; this strategy is analog to ground-based transmission spectroscopy, where the broadband spectra are normalized against the continuum flux variations caused by the Earth's atmosphere, but narrow-band absorption signals are conserved. We identified that the white-light absolute flux of WASP-76 during the fourth orbit of visit A is consistent with that of the fourth orbit of visit D. Thus, we assume that the star was well-centered in the slit during this orbit and apply a multiplicative correction factor in each spectrum of the time-series to normalize them to the fourth orbit white-light flux. After normalizing the spectra in visit A, we proceeded to calculate a Mg\,{\sc ii} light curve similarly to visit D. We then co-add the normalized light curves and do not fit an additional baseline offset. The best-fit models yield $R_{\rm eff}$(Mg\,{\sc ii}, k)$/R_s$ = 0.3710 $\pm$ 0.0356 and $R_{\rm eff}$(Mg\,{\sc ii}, h)$/R_s$ = 0.3849 $\pm$ 0.0326 (see \autoref{fig:6}). For reference, \autoref{fig:6} also shows the optical continuum radius from \citet{Fu_2021} and the Roche-lobe radius scale. We note that some data points at late phases remain below unity in the co-added light curve, which is suggestive of post-transit absorption, but we do not attempt to model this feature with the current dataset.

Subsequently, we investigated the Fe lines; Fe\,{\sc i} exhibit ground-state transitions at central wavelengths of 2484, 2523, 2719, 2913 and 2937 {\AA}. As we see in \autoref{fig:12}, we do not see any obvious absorption appear in the data in any of these transitions. The non-detection of Fe I is not surprising: with an ionization potential of $\sim$ 7.9 eV, essentially identical to that of Mg ($\sim$ 7.6 eV), Fe should be efficiently photoionized in the upper atmosphere of WASP-76b under the same irradiation conditions that produce the observed Mg\, {\sc ii} signal. Following \autoref{sec: Results - narrowband}, the narrow-band transmission spectrum showed an increased $R_p/R_s$ of 0.1828$_{-0.0780}^{+0.1396}$ for GP and 0.1938$_{-0.0717}^{+0.1264}$ for linear model at 2383.5 {\AA}, which could correspond to a transition of Fe\,{\sc ii}. Fe\,{\sc ii} show transitions in series of multiplets (as seen in \autoref{fig:1}) with wavelengths of 2600, 2382 and 2344 {\AA} for the UV1, UV2 and UV3 multiplets respectively. However, when we look at the in-transit and out-of-transit spectra for each of these lines, we do not see any obvious absorption (see \autoref{fig:13}). We discuss potential reasons why we do not see the absorption in \autoref{sec: Discussion - ionization}.

\subsubsection{Mg\,{\sc ii} spectrophotometry by single line analysis} \label{sec: Results - narrower bins}

\begin{figure}[h]
    \includegraphics[width=0.48\textwidth]{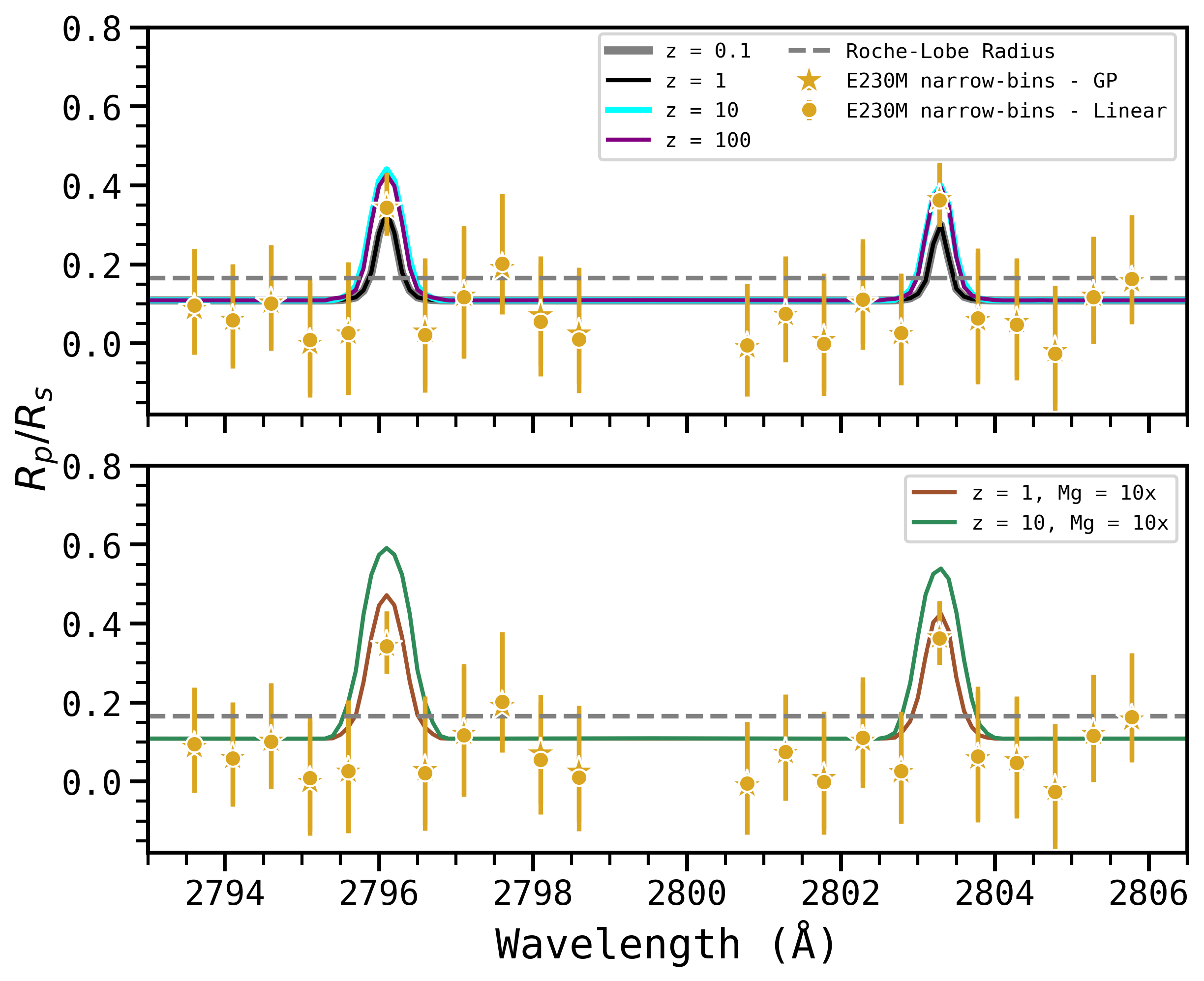}
    \caption{Targeted spectrophotometry around the Mg\,{\sc ii} doublet for visit D. Points show the transmission spectrum derived from 0.5 {\AA} bins centered on the blue wing of each line (five bins on either side). Golden symbols indicate measurements using the GP detrending (stars) and linear detrending (circles). The horizontal dashed lines indicate the Roche-lobe scale. Solid curves show several \texttt{sunbather} forward models for qualitative comparison only; a quantitative model comparison is presented in \autoref{sec: model comparison}. \textit{Top:} models varying bulk metallicity. \textit{Bottom:} models varying metallicity and Mg abundance.}
    \label{fig:7}
\end{figure}

Because the Mg\,{\sc ii} absorption is localized in the blue wing of the doublet, we next construct targeted sub-{\AA} spectrophotometric bins centered on that region and measure a corresponding $R_{\rm eff}/R_s$ using the same methodology as described in the spectroscopic fits' section (\autoref{sec:spectroscopic fits}). It was noticeable that the binning widths used mattered with respect to the strength of the signal. We tested a range of bin widths while monitoring the amounts of pixels covered within the bin so we would not be just counting noise. The optimal bin width when inspecting the Mg\,{\sc ii} doublet was 0.5 {\AA}, which covers 11 pixels. The data was binned around both lines, generating five bins on each side of the line, the center of each line was chosen based on the information gathered from \autoref{sec: Results - high res} which is on the blue wing of the line and corresponds to 2796.1 {\AA} for $k$ and 2803.28 {\AA} for $h$. The resulting transmission spectrum is shown on \autoref{fig:7}. We show the results for both lines using GPs and the linear model for systematic corrections. For the GP model, $R_{\rm eff}$(Mg\,{\sc ii}, k)/$R_s$ = 0.3435$_{-0.0696}^{+0.0882}$ and $R_{\rm eff}$(Mg\,{\sc ii}, h)/$R_s$ = 0.3653$_{-0.0664}^{+0.0869}$ and for the linear model $R_{\rm eff}$(Mg\,{\sc ii}, k)/$R_s$ = 0.3433$_{-0.0699}^{+0.0874}$ and $R_{\rm eff}$(Mg\,{\sc ii}, h)/$R_s$ = 0.3631$_{-0.0675}^{+0.0943}$. Similarly as in \autoref{sec: Results - high res}, we include the continuum and the Roche radius for comparison and we notice that the Mg\,{\sc ii} radius from each component exceeds those values, once again. \autoref{tab: radius_comparison} summarizes the comparison between the Mg\,{\sc ii} effective radii, the optical radius and the Roche-lobe scale. We additionally note that the choice of systematics model (GP vs linear) has little effect on the inferred $R_{\rm eff}/R_s$ in these targeted 0.5 {\AA} bins (\autoref{fig:14}), in contrast to the white light curve fits (see \autoref{fig:2}). Both detrending approaches yield consistent $R_{\rm eff}/R_s$ values at the achieved precision (\autoref{Appendix: Mg II}). For visual context, we overplot a small set of \texttt{sunbather} \citep{Linssen_2024} forward models (more details on the models in \autoref{sec: upper atmo}). A quantitative model comparison and discussion is presented in \autoref{sec: model comparison}.

\begin{table}[ht]
    \centering
    \caption{Reference radius ratios and Mg\,{\sc ii} effective radii measured in this work. Optical $R_p/R_s$ from \citet{Fu_2021}; Roche-lobe computed in this work using the \citet{Eggleton_1983} approximation.}
    \begin{tabular}{l c c}
    \hline\hline
    \multicolumn{3}{c}{Reference scales} \\
    \hline
    Optical $R_p/R_s$ & & $0.1087\pm0.0005$ \\
    Roche-lobe $R_{\rm RL}/R_s$ & & $0.1659^{+0.0041}_{-0.0042}$\\
    \hline
    \multicolumn{3}{c}{Mg\,{\sc ii} effective radii ($R_{\rm eff}/R_s$)}\\
    \hline
    & Mg\,{\sc ii} $k$ & Mg\,{\sc ii} $h$ \\
    single line analysis & $0.3710\pm0.0356$
 & $0.3849\pm0.0326$\\
    0.5 {\AA} bins, GP & $0.3435^{+0.0882}_{-0.0696}$ & $0.3653^{+0.0869}_{-0.0664}$\\
    0.5 {\AA} bins, linear & $0.3433^{+0.0874}_{-0.0699}$ & $0.3631^{+0.0943}_{-0.0675}$\\
    \hline
    \end{tabular}
    \label{tab: radius_comparison}
\end{table}

\section{Atmospheric Modeling} \label{sec:Modeling}

In the following section we model WASP-76b's atmosphere on two complementary scales. First, we perform atmospheric retrievals on the lower atmosphere focused on the continuum and molecular absorbers. Second, we model the upper atmosphere with forward hydrodynamic/photoionization calculations to interpret the Mg\,{\sc ii} resonance-line absorption and its implications for high-altitude escape.

\subsection{Lower-atmosphere modeling} \label{sec: low atmo}

We perform a retrieval analysis to test whether adding NUV broadband information improves constraints on SiO in WASP-76b. This motivation follows \citet{Lothringer_2022}, who showed for the UHJ WASP-178b the short-wavelength slope can be produced either by atomic metal opacity or molecular absorption from species such as SiO, leading to a degeneracy in broadband interpretations. In WASP-76b, our high-resolution analysis reveals Mg\,{\sc ii} absorption in the NUV, demonstrating that metals are present at high altitudes; however, this does not by itself rule out additional SiO opacity deeper in the atmosphere.

We used the \texttt{petitRADTRANS} retrieval framework \citep{Molliere_2019,Nasedkin_2024} assuming a cloud-free atmosphere with 80 log-spaced layers between $10^{-6}$ and $10^3$ bar. We fit for free  molecular abundances of H$_2$O \citep{Rothman_2010}, FeH \citep{Wende_2010}, SiO \citep{Yurchenko_2022}, TiO \citep{McKemmish_2019}, and VO \citep{McKemmish_2016}, as well as atomic abundances of Fe, Fe+, Mg, and Mg+ \citep{Kurucz_1995}. All abundances assumed uniform priors between 10$^{-10}$ and $1.0$. We used the \cite{Guillot_2010} temperature structure parameterization. We ran the retrieval inside a Multinest nested sampling statistical framework \citep{Feroz_2008,Feroz_2009} with 2000 live points.

\begin{figure}[h]
    \includegraphics[width=0.5 \textwidth]{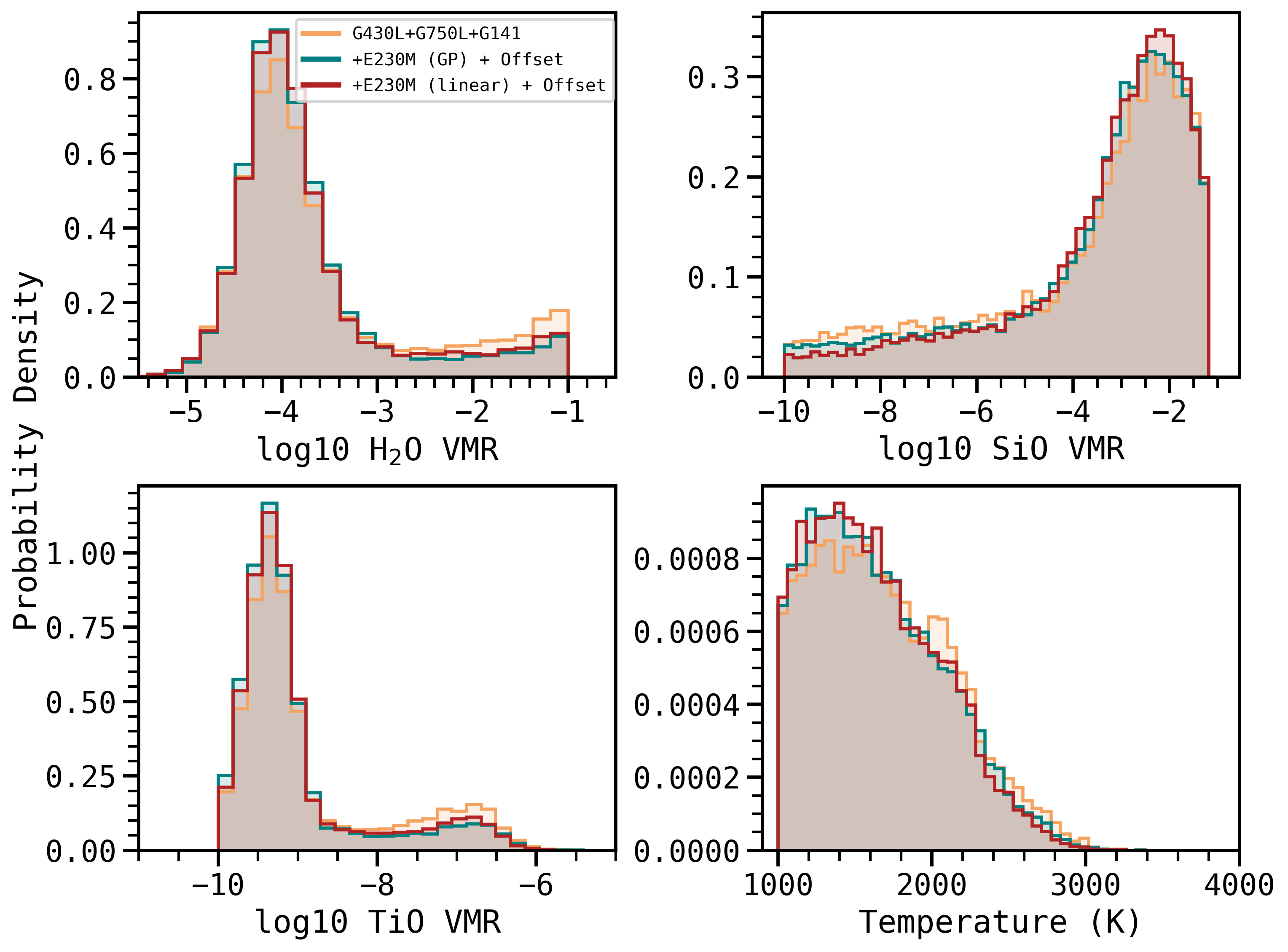}
    \caption{Posterior distributions from \texttt{petitRADTRANS}-based retrievals of the broadband transmission spectrum. Sand shows the optical+IR dataset (STIS/G430L + STIS/G750L + WFC3/G141; \citealt{Fu_2021}). Teal and burgundy show retrievals including the STIS/E230M broadband points, using two methods for systematics (GP + offset and linear + offset, respectively). The posteriors show the volume mixing ratios (VMRs) of H$_2$O, SiO and TiO, and the atmospheric temperature. The results are broadly consistent across the three dataset treatments.}
    \label{fig:8}
\end{figure}

We  first fit the broadband transmission spectrum (STIS/G430L + STIS/G750L + WFC3/G141) from \cite{Fu_2021} and then repeated the fits with the inclusion of the new STIS/E230M broadband points, testing two systematics treatments for E230M (a GP + offset and a linear model + offset). The posteriors shown in \autoref{fig:8} indicate that adding the E230M does not qualitatively change the inferred picture: H$_2$O is robustly constrained by the IR data, while TiO is not detected. The atmospheric temperature is moderately constrained, with posterior probability in the $\sim$ 1,200 $-$ 2,200 K range but a long high-T tail, and SiO remains only weakly constrained. The three dataset treatments, produce very similar posteriors overall, indicating that the NUV data and the choice of systematics model slightly change but do not overturn the broadband retrieval results. We attribute the poor constraining power of the E230M data not only to the large broadband errorbars, but also to the offsets that were fit for between the NUV and optical data due to the poorly constrained absolute transit depth. This points toward the importance of the UV-optical coverage available with WFC3/G280 \citep[e.g.,][]{Wakeford_2020}, in synergy with the high-resolution NUV STIS/E230M observations that are sensitive to atmospheric escape \citep[e.g.,][]{Lothringer_2022}.

\subsection{Upper-atmosphere modeling} \label{sec: upper atmo}

The upper atmosphere of WASP-76b is probed by specific spectral lines that are formed at those layers, such as the Mg\,{\sc i} and Mg\,{\sc ii} lines in the NUV. In order to interpret the excess in-transit absorption described in \autoref{sec: Results - high res}, we utilize the {\tt sunbather} Python code \citep{Linssen_2024}. This framework employs a Parker-wind \citep{Parker_1958} approximation to solve the advection and ionization balance of the upper atmosphere of a H/He-dominated exoplanet \citep{Oklopcic&Hirata_2018, p-winds} in combination with the Cloudy photoionization code \citep{Ferland_1998, Ferland_2017, Chatzikos_2023} to calculate the population of specific species in the upper atmosphere. 

The \texttt{sunbather} models take as input the planetary parameters, stellar parameters, stellar high-energy SED, mass-loss rate, isothermal temperature, metallicity and abundance of Mg. We took the system parameters from \citet{Fu_2021}. Since the X-rays to extreme UV (XUV) spectrum of WASP-76 is not determined, we assume the high-energy SED of WASP-17 from the MUSCLES Extension program \citep{Behr_2023} as a proxy. We estimated the mass-loss rate and upper-atmosphere temperature using the ATES code \citep{Caldiroli_2021, Caldiroli_2022} obtaining $\dot{M}\approx10^{12.14}\,\mathrm{g\,s^{-1}}$; for the latter, we assume that the isothermal value corresponds to the maximum of the temperature profile calculated by ATES\footnote{\footnotesize{This is the same approach used by \citet{p-winds} and \citet{Lampon_2020} to match results from the isothermal Parker wind model with a more self-consistent hydrodynamic model.}}. In the remainder of this section we parameterize the bulk metallicity as a scaling relative to solar, $z \equiv Z/Z_\odot$, where $Z$ denotes the heavy-element abundance (elements heavier than He) in the assumed composition. Thus $z=1$ corresponds to solar metallicity, $z=0.1$ to sub-solar metallicity, and $z=10$–100 to super-solar metallicities. We calculate transmission spectra with \texttt{sunbather} for a range of chemical compositions: 

\begin{enumerate}[label=(\alph*)]
    \item $z=0.1\times$ solar metallicity,
    \item solar metallicity ($z=1$),
    \item $z=10 \times$ solar metallicity,
    \item $z=100 \times$ solar metallicity,
    \item solar metallicity ($z=1$) and $10 \times$ solar Mg abundance,
    \item $z=10 \times$ solar metallicity and $10 \times$ solar Mg abundance.
\end{enumerate}

\begin{figure}[h]
    \includegraphics[width=0.48 \textwidth]{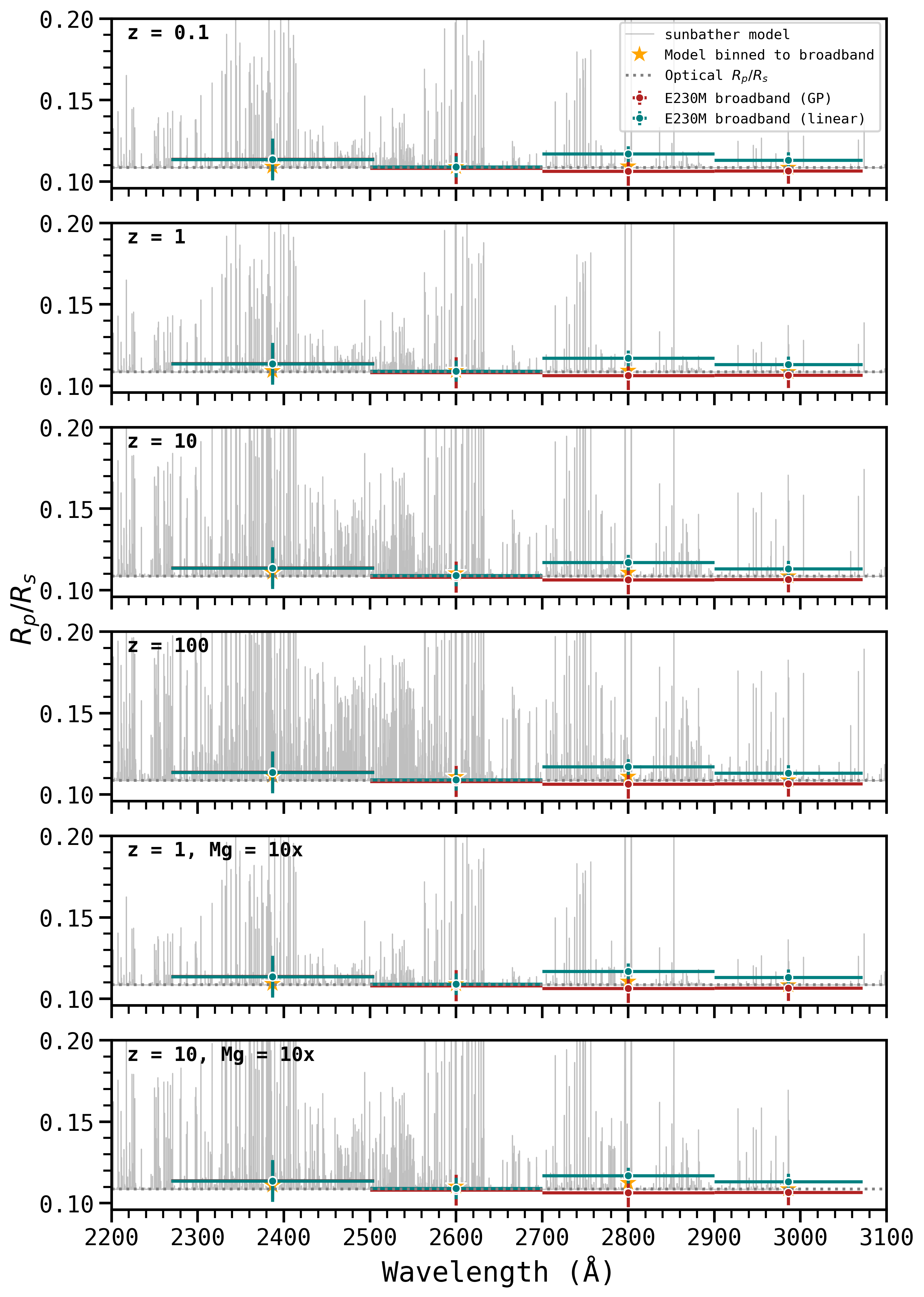}
    \caption{Comparison between the STIS/E230M broadband transmission spectrum and an upper-atmosphere \texttt{sunbather} model. Colored points show the measured radius ratios for visit D with the different systematics treatment (GP in burgundy and linear in teal) and orange shows the model binned to the same wavelength channels as the broadband points. In gray we show different high resolution models computed with a range of chemical compositions: sub-solar, solar composition, $10\times$ solar metallicity, $100\times$ solar metallicity, solar metallicity and $10\times$ solar Mg abundance and, $10\times$ solar metallicity and $10\times$ solar Mg abundance. The dotted line corresponds to the optical radius.}
    \label{fig:9}
\end{figure}

\subsection{Model-data comparison and results} \label{sec: model comparison}

We compare the \texttt{sunbather} model grid to two observables: (i) the STIS/E230M broadband transmission spectrum (\autoref{sec: Results- broadband}) and (ii) the targeted 0.5 {\AA} Mg\,{\sc ii} transmission spectrum of the $k$ and $h$ lines (\autoref{sec: Results - narrower bins}). For the broadband comparison, we bin each model spectrum to the same wavelength bins as the E230M broadband points. For the targeted Mg\,{\sc ii} comparison, we interpolate the model to the same 0.5 {\AA} bins used for the data. 

For each dataset and systematics treatment (GP and linear), we compare the forward models using the chi-square statistic assuming Gaussian errors,
\begin{equation}
\chi^{2} = \sum\limits_{i=1}^{N}\left(\frac {y_i - \mu_i} {\sigma_i}\right)^2,
\label{chi2} 
\end{equation}
where $y_i$ and $\sigma_i$ are the measured values and uncertainties, and $\mu_i$ is the model prediction evaluated at the same wavelength bin. Because we are treating the \texttt{sunbather} spectra as forward models (no parameters were fitted, $k$ = 0), the degrees of freedom (DOF) are $\nu = N$.

To rank models within a given subset (fixed dataset and systematics treatment), we use the Akaike Information Criterion (AIC: \citealt{Akaike_1974}) following \citet{Thorngren_2026}. For Gaussian errors with fixed $\sigma_i$, the log-likelihood is $\ln L = \mathrm{const}-\chi^2/2$,
\begin{equation}
\mathrm{AIC} = -2\ln L + 2k \simeq \chi^2 + 2k.
\label{eq:aic}
\end{equation}
With $k=0$ for all models in our grid, $\mathrm{AIC} \simeq\chi^2$, and model ranking by AIC is equivalent to ranking by $\chi^2$. We report $\Delta\mathrm{AIC} = \mathrm{AIC} - \mathrm{AIC}_{\min}$ within each subset and convert these to AIC model probabilities (Akaike weights) as
\begin{equation}
p(M_j)\equiv p_{\mathrm{AIC}}(M_j\mid y)=
\frac{\exp\!\left(-\frac{1}{2}\Delta\mathrm{AIC}_j\right)}
{\sum\limits_{m}\exp\!\left(-\frac{1}{2}\Delta\mathrm{AIC}_m\right)}.
\label{eq:pAIC}
\end{equation}
We interpret $p(M_j)$ as the probability of model $M_j$ within the tested model set under equal prior weights across models. For compact presentation we also report the complementary probability $p(\neg M_j)=1-p(M_j)$, which we refer to as the ``disfavored probability within the model set.''
 
\autoref{fig:9} shows the broadband E230M spectrum compared to the model grid and the targeted Mg\,{\sc ii} measurements in \autoref{fig:7}. \autoref{tab: model_comparison} summarizes the comparison between the \texttt{sunbather} grid and the combined dataset consisting of the E230M broadband points (N = 4) together with the targeted Mg\,{\sc ii} 0.5 {\AA} bins (N = 22), for both systematics treatments. The broadband points alone provide weak discrimination among compositions, whereas the Mg\,{\sc ii} bins dominate the total $\chi^2$ and drive the model ranking (see \autoref{tab: model_comparison_indiv} for individual comparison). Under the GP systematics treatment, the solar-composition model ($z=1$) yields the minimum AIC, with the sub-solar ($z=0.1$), $z=10$, and $z=100$ models remaining competitive ($\Delta\mathrm{AIC}\lesssim 1$). Under the linear systematics treatment, the best-fitting model shifts slightly to $z=100$, but $z=1$, $z=0.1$, and $z=10$ remain close in AIC ($\Delta\mathrm{AIC}\lesssim 0.6$), indicating that the data are consistent with metallicities ranging from sub-solar to $\sim$ 100$\times$ solar within the tested grid. In contrast, models with extreme Mg enhancement are disfavored in both systematics treatments: the $z=10$ with Mg$\times 10$ model is strongly disfavored ($\Delta\mathrm{AIC}\gtrsim 15$), and the $z=1$ with Mg$\times 10$ model is moderately disfavored ($\Delta\mathrm{AIC}\approx 2.6$--2.9). We therefore adopt the conservative interpretation that the combined data are consistent with metallicities spanning sub-solar to $100\times$ solar within the tested grid, while strongly disfavoring extreme Mg enhancement. The associated AIC model probabilities 
(\autoref{eq:pAIC}) are conditional on the finite tested model set and equal prior weights.

\begin{table*}[t]
\centering
\caption{\texttt{sunbather} model comparison to the combined dataset (N = 26) consisting of N = 4 STIS/E230M broadband points and N = 22 Mg\,{\sc ii} targeted bins for GP and linear systematics treatment. We report $\chi^2$, AIC, $\Delta$AIC (relative to the best model within each subset), and AIC model probabilities $p(M)$ normalized within each subset. We also report $p(\neg M)=1-p(M)$ as the disfavored probability within the tested model set.}
\begin{tabular}{l l r r r r r}
\hline
\hline
Systematics & Model & $\chi^2$ & AIC & $\Delta$AIC & $p(M)$ (\%) & $p(\neg M)$ (\%)\\
\hline
\multicolumn{7}{c}{Combined data (N = 26)}\\
\hline
GP & $z=1$ & 6.00 & 6.00 & 0.00 & 28.3 & 71.7\\
GP     & $z=0.1$ & 6.05 & 6.05 & 0.05 & 27.5 & 72.5\\
GP     & $z=100$ & 6.62 & 6.62 & 0.62 & 20.7 & 79.3\\
GP     & $z=10$ & 7.04 & 7.04 & 1.04 & 16.8 & 83.2\\
GP     & $z=1$, Mg$\times$10 & 8.86 & 8.86 & 2.86 & 6.76 & 93.2\\
GP     & $z=10$, Mg$\times$10 & 22.8 & 22.8 & 16.8 & 0.01 & 99.9\\

\hline
Linear & $z=100$ & 8.77 & 8.77 & 0.00 & 27.4 & 72.6\\
Linear & $z=1$ & 9.15 & 9.15 & 0.38 & 22.7 & 77.3\\
Linear & $z=0.1$ & 9.21 & 9.21 & 0.44 & 22.0 & 78.0 \\
Linear & $z=10$ & 9.37 & 9.37 & 0.60 & 20.4 & 79.6\\
Linear & $z=1$, Mg$\times$10 & 11.4 & 11.4 & 2.59 & 7.51 & 92.5\\
Linear & $z=10$, Mg$\times$10 & 23.7 & 23.7 & 14.9 & 0.02 & 99.9\\
\hline
\end{tabular}
\label{tab: model_comparison}
\end{table*}

\section{Discussion} \label{sec:Discussion}

\subsection{Mg\,{\sc ii} absorption: kinematics and observational sensitivity}

We detect significant excess absorption in the Mg\,{\sc ii} doublet at 2796.35 {\AA} ($k$) and 2803.53 {\AA} ($h$) in the transmission spectrum of WASP-76b. Using the single line analysis, we measure an in-transit absorption at the blue wing of each line. We also recover consistent absorption using targeted 0.5 {\AA} spectrophotometric bins. Parameterizing the line-band absorption depth with an effective radius ratio yields consistent values for both components across the two analyses: the single line analysis gives $R_{\rm eff}$(Mg\,{\sc ii}, k)$/R_s$ = 0.3710 $\pm$ 0.0356 and $R_{\rm eff}$(Mg\,{\sc ii}, h)$/R_s$ = 0.3849 $\pm$ 0.0326, while the targeted 0.5 {\AA} spectrophotometry yields $R_{\rm eff}/R_s$ $\sim$ 0.34 $-$ 0.37 with larger uncertainties (\autoref{sec: Results - narrower bins}). This detection indicates that Mg\,{\sc ii} is present at high altitudes in the atmosphere of WASP-76b and can be probed with NUV spectroscopy.

The interpretation of an absorption at negative velocities ([-50, -10] km s$^{-1}$) in the stellar rest frame implies material with a line-of-sight motion towards the observer during transit. In this context, such blue-shifted absorption can arise from a combination of hydrodynamic expansion/escape \citep[e.g.,][]{Yelle_2004,Murray-Clay_2009,Owen_2019}, strong winds near the terminator that project toward the observer \citep[e.g.,][]{Showman&Guillot_2002,Ehrenreich_2020,Seidel_2025}, radiative acceleration acting on resonance-line ions \citep[e.g.,][]{Vidal-Madjar_2003,Bourrier&Lecavelier_2013}, interactions with the stellar wind \citep[e.g.,][]{Bourrier_2016, Carolan_2021}, and tidal deformation \citep[e.g.,][]{Owen_2023, Allart_2025}. While our current data does not constrain the geometry of the absorbing gas, the detection of Mg\,{\sc ii} at velocities over [-50, -10] km s$^{-1}$ suggests a fast, irradiated, and extended ionized layer in the upper atmosphere.

A key observational limitation in Mg\,{\sc ii} is the deep stellar line core: the stellar flux approaches very low values near the center of the line, which strongly reduces sensitivity to planetary absorption near $v\approx$ 0 km s$^{-1}$. This means that our measurements of Mg\,{\sc ii} provide constraints in the wings of the line, where the stellar flux is sufficient to detect additional absorption. Having this ``core masking", means that our wings-only signature should not be interpreted as the core of the planetary Mg\,{\sc ii} being blue-shifted or as evidence that low-velocity Mg\,{\sc ii} is absent; rather it indicates that the data preferentially reveal a high-altitude component that imprints absorption in the wings. We also considered whether stellar Mg\,{\sc ii} line variability could produce the apparent in-transit excess. Comparing the integrated out-of-transit Mg\,{\sc ii} flux between visit A (January 2022) and visit D (February 2023), separated by over a year, we find consistent flux levels within measurement uncertainties, arguing against large-amplitude epoch-to-epoch variability in the stellar line. Nevertheless, we cannot completely rule out variability on shorter timescales (within a visit) or at low amplitude with the current dataset. F-type stars are generally considered less chromospherically active than later-type stars \citep{Chahal_2023}, but NUV variability in F stars is not well characterized.

\subsection{Roche-lobe and spatial extent of the absorbing gas} \label{sec: Discussion - Roche}

To place our NUV measurements in the context of atmospheric escape, we compare two physically well-defined size scales -- optical transit radius and the Roche-lobe radius -- and use the Mg\,{\sc ii} absorption as a diagnostic of how far above the optical photosphere/continuum radius ionized material may extend. We adopt the optical radius from \citet{Fu_2021}, $R_p/R_s$ = 0.1087$\pm$0.0005, as the baseline planetary size traced by continuum opacity. We estimate the planet's Roche-lobe radius using the Eggleton approximation \citep{Eggleton_1983}, 

\begin{equation}
 \frac{R_{\rm RL}}{a} = \frac{0.49\,q^{2/3}}{0.6\,q^{2/3}+\ln\!\left(1+q^{1/3}\right)},   
\end{equation}

where $q=M_{\rm p}/M_s$ and $a$ is the orbital separation. We convert this to a stellar-normalized Roche-lobe radius using the scaled semi-major axis,

\begin{equation}
    \frac{R_{\rm RL}}{R_s}= \left(\frac{R_{\rm RL}}{a}\right)\left(\frac{a}{R_s}\right).
\end{equation}

We used the system parameters listed in \autoref{tab: sys_params} to obtain $R_{\rm RL}/R_s$ = 0.1659$_{-0.0042}^{+0.0041}$ for WASP-76b. See \citet{Erkaev_2007} for discussion of how Roche geometry modifies atmospheric escape estimates. As a visual reference, we overplot in \autoref{fig:6} and \autoref{fig:7} a simple ``Roche-lobe transit'' curve corresponding to a gray occulting disk with radius fixed to $R_{\rm RL}/R_s$; this curve is intended as an illustrative benchmark rather than a physical model for resonance-line absorption.
 
With this reference value we can interpret the apparent ``large-radius'' absorption in the narrow-band transmission spectrum. In the 4 {\AA} narrow-band spectrum (\autoref{fig:4}), some isolated bins towards the blue yield $R_p/R_s$ values exceeding $R_{\rm RL}/R_s$; however, these points are not associated with the expected wavelengths of strong metal transitions. We therefore do not interpret the 4 {\AA} scan alone as evidence that the atmosphere overfills the Roche-lobe.

The Mg\,{\sc ii} detection provides a more physically motivated probe of high-altitude gas. In both our single line analysis (\autoref{fig:6}) and the targeted 0.5 {\AA} bins (\autoref{fig:7}), the inferred Mg\,{\sc ii} absorption corresponds to an effective occulting area exceeding the optical continuum and the Roche-lobe scale ($R_{\rm eff}/R_s > R_{\rm RL}/R_s$), implying the presence of Mg\,{\sc ii} gas beyond the planet's Roche-lobe. Additional NUV transits with higher SNR and improved phase coverage will be needed to robustly characterize the escaping geometry and test for signatures such as early ingress or late egress.

\subsection{Ionization/composition constraints from non-detections (Mg\,{\sc i}, Fe\,{\sc i} and Fe\,{\sc ii})} \label{sec: Discussion - ionization}

Combining the Mg\,{\sc ii} detection with the lack of Mg\,{\sc i}, qualitatively supports a highly ionized upper atmosphere in WASP-76b. UHJs undergo intense irradiation and photoionization is efficient at low pressures, making  Mg\,{\sc ii} a more sensitive tracer of the highest-altitude layers than neutral Mg\,{\sc i} \citep[e.g.,][]{Yelle_2004,GarciaMunoz_2007,Murray-Clay_2009,Lothringer_2018,Parmentier_2018,Owen_2019}. In this scenario, the non-detection of Mg\,{\sc i} does not contradict the presence of metals at high-altitude, rather, it supports that Mg predominantly ionized where the NUV resonance lines form.

We do not detect absorption in the Fe\,{\sc i} transitions nor in the Fe\,{\sc ii} multiplets in the NUV. This could be a combination of limited sensitivity (low flux in the NUV and line blending) and atmospheric variability or spatial structure \citep[e.g.,][]{Sing_2019,Bourrier&Lecavelier_2013}. More specifically, the optical Fe\,{\sc i} absorption in WASP-76b was reported to be phase dependent and asymmetric \citep{Ehrenreich_2020}, potentially linked to condensation/rainout and strong dynamics; if the similar behavior affects NUV iron lines, co-adding spectra across phases and visits could dilute a localized signal. 

The nature of a strong Mg\,{\sc ii} detection with non-detections of Mg\,{\sc i} and the iron lines is therefore consistent with a highly ionized and dynamically complex upper atmosphere where different species (and ionization stages) may trace different regions and/or velocities of the flow. We require additional NUV transits with higher SNR and phase-resolved analysis to place tighter constraints on neutral and ionized iron to test whether the Mg\,{\sc ii} absorption exhibits ingress/egress asymmetries analogous to those seen in the optical \citep{Ehrenreich_2020}.

\subsection{Context of the observation and models}

A useful comparison for our NUV results is WASP-121b, which was observed with HST/STIS E230M over a comparable wavelength range and resolving power \citep{Sing_2019}. That study reports a strong NUV transit depth excess relative to the optical/near-IR continuum and detects spectrally resolved exospheric absorption in both components of the Mg\,{\sc ii} doublet (as well as Fe\,{\sc ii}), with  $R_{p}/R_s$ exceeding the Roche-lobe scale in transit geometry. Recent 3D hydrodynamic simulations of WASP-121b by \citet{Wang_2026} provide a physical framework for interpreting such UV metal signatures, predicting that stellar irradiation can drive outflows that are shaped into extended structures by orbital motion and stellar gravity. In these models, different tracers probe distinct regions of the flow and can exhibit line-of-sight velocities of order tens of km s$^{-1}$, qualitatively consistent with the high-velocity components reported in the transmission spectra of WASP‑121b (e.g., \citealt{Seidel_2025}). This highlights that UV observations constrain not only composition but also the geometry and kinematics of the upper atmosphere. For WASP-76b we also resolve both Mg\,{\sc ii} components and infer effective radii that exceed the Roche-lobe benchmark (\autoref{sec: Discussion - Roche}), consistent with a highly extended ionized component. In contrast to WASP-121b, our E230M broadband transmission points show no statistically significant wavelength-dependent variations at the achieved precision; the Mg\,{\sc ii} signal in WASP-76b is most clearly detected as velocity-localized absorption in the line wings and recovered with targeted 0.5 {\AA} bins. Together, these results suggest that while Mg\,{\sc ii} absorption at large effective radii is observed in multiple UHJs accessible in the NUV, the detectability of a broadband NUV continuum excess and additional metal species (e.g. Fe\,{\sc ii}) may depend sensitively on system parameters and data quality.

Broadly placing WASP-76b alongside other hot Jupiters observed with STIS/E230M highlights both commonalities and diversity in NUV escape diagnostics. WASP-178b shows no clear Mg\,{\sc ii} absorption in E230M \citep{Lothringer_2022}, WASP-79b shows a non-detection on Mg\,{\sc ii} (or any metals) \citep{Gressier_2023} while KELT-9b shows strong Mg\,{\sc ii} and Fe\,{\sc ii} signatures consistent with rapid mass-loss and substantial line broadening \citep{Baldwin_2026}. In this emerging E230M sample \citep{Sing_2019,Lothringer_2022,Gressier_2023,Baldwin_2026}, WASP-76b occupies an intermediate regime: we detect Mg\,{\sc ii} in both doublet components but no Fe\,{\sc ii} and do not measure significant broadband NUV excess. \citet{Baldwin_2026} proposed an empirical Mg\,{\sc ii} Cosmic Shoreline in which the detection of escaping Mg\,{\sc ii} is governed primarily by the planet's escape velocity and the incident XUV flux: planets with high XUV irradiation and relatively low $V_{esc}$ are most likely to show observable Mg\,{\sc ii} outflows. WASP-76b lies in the region of parameter space associated with Mg\,{\sc ii} escape (above the non-detection of WASP-79b and below the extreme KELT-9b), so our detection provides an intermediate data point that helps refine the shoreline boundary between planets with and without Mg\,{\sc ii} escape. 

From a modeling standpoint, our results emphasize that the STIS/E230M dataset constrains the upper and lower atmosphere in qualitatively different ways. The broadband E230M points carry limited information on continuum opacities because their uncertainties are large and because inter-instrument offsets absorb much of the leverage on the short-wavelength slope; consequently, adding E230M does not break the optical/IR degeneracies in the lower-atmosphere retrieval (e.g., between metal opacity and SiO) and the inferred composition is still driven primarily by the optical+IR data. Consistently with this, our AIC-based model comparison shows that the broadband spectrum does not provide meaningful distinctions among the tested compositions (\autoref{tab: model_comparison}).

By contrast, the spectrally resolved Mg\,{\sc ii} measurement provides stronger discrimination within the same model grid because they require an extended Mg\,{\sc ii} column in the line wings. In our comparison, models with extreme Mg enhancement ($10\times$ metallicity and $10\times$ solar Mg abundance) over-produce the absorption in the observable wings and are strongly disfavored by the Mg\,{\sc ii} bins ($\Delta$AIC $\simeq$ 16; $p(M)\lesssim0.01\%$ within the tested model set), whereas models with moderate metallicities (0.1$\times$ $-$ 100$\times$ solar) remain competitive ($\Delta$AIC $\lesssim 1$--2; see \autoref{fig:7}). Notably, this upper-atmosphere inference is broadly consistent with the near-solar metallicity inferred from optical+IR spectra \citep{Fu_2021}: the Mg\,{\sc ii} constraints do not require strong Mg enrichment or extreme bulk metallicities. While our data does not directly measure elemental abundances in the escaping flow, the fact that the Mg\,{\sc ii} wing absorption can be reproduced without invoking extreme Mg enrichment suggests that a strongly Mg-fractionated outflow is not required. This qualitative consistency is compatible with a hydrodynamic escape scenario in which the outflow is collisional over much of the upper atmosphere, so that heavy species can be dragged along with the bulk gas; however, ionization, condensation/rainout, and radiative forces can still modify the observable ionic composition with altitude.

While these 1D models capture the gross Mg\,{\sc ii} absorption level, interpreting the signal in terms of the spatial extent and mass loss ultimately requires modeling beyond a gray-occultation picture. Resonance-line measurements are sensitive to the optical depth, covering fraction, and velocity field of the absorbing gas, and the stellar Mg\,{\sc ii} line profile limits sensitivity near line center. Forward models that couple hydrodynamic structure with ionization balance and radiative forces (and ultimately 3D geometry) are therefore needed to connect the observed Mg\,{\sc ii} line profiles and light curves to physical constraints on atmospheric escape and Roche-lobe scale structure \citep[e.g.,][]{Bourrier&Lecavelier_2013,Shaikhislamov_2018,Carolan_2021,Presa_2024,Vidotto_2025}.

Upper-atmosphere models are also increasingly constrained by combining multiple tracers. For WASP-76b, He\,{\sc i} 10830 {\AA} measurements provide an upper limit (or weak/tentative constraint) on metastable helium absorption and corresponding bounds on the mass-loss rate \citep{Lampon_2023}. Jointly interpreting He\,{\sc i} and Mg\,{\sc ii} is valuable because these diagnostics probe different thermochemical regimes and can help discriminate between models that reproduce a strong metal signature but over-predict the helium signal (or vice versa). In this sense, future multi-wavelength analyses that treat He\,{\sc i} and UV metal lines within a consistent escape framework will be particularly informative for constraining $\dot{M}$ and the ionization structure of the outflow.

\section{Conclusions} \label{sec:Conclusion}

We presented NUV observations of the UHJ WASP-76b obtained with HST/STIS E230M, covering 2275 $-$ 3119 {\AA}. Using a spectrophotometric approach (broadband $\sim$ 200 {\AA} bins) and a narrow-band of 4 {\AA} bins, we do not find strong evidence for metal absorption across the full wavelength range.

In contrast, a single line analysis reveals significant excess absorption in the Mg\,{\sc ii} doublet at 2796.35 {\AA} ($k$) and 2803.53 {\AA} ($h$). The absorption is found at the blue wing over velocities [-50, -10] km s$^{-1}$ and is detected in both components. Assisted by the velocity localization, we also recover consistent absorption in targeted 0.5 {\AA} spectrophotometric bins centered on the wing. In both the velocity-window line light curves and the targeted 0.5 {\AA} bins, the inferred effective radii for each Mg\,{\sc ii} component exceed the Roche-lobe scale ($R_{\rm eff}/R_s > R_{\rm RL}/R_s$), suggesting absorption from a highly extended Mg\,{\sc ii} component consistent with Roche-lobe overflow. A robust demonstration of Roche‑lobe overflow will require additional NUV transit observations with improved phase sampling of the Mg\,{\sc ii} absorption (to test for early ingress/late egress) and forward modeling of the line profiles.

We do not detect absorption in Mg\,{\sc i}, Fe\,{\sc i}, or Fe\,{\sc ii} in the lines examined, which may reflect a combination of limited sensitivity in the NUV and a highly ionized, spatially/temporally structured upper atmosphere. We also performed preliminary broadband retrievals to test whether adding the STIS/E230M NUV points improves constraints on SiO, motivated by the low-resolution degeneracy between metal opacity and SiO highlighted by \citet{Lothringer_2022}; at the achieved precision, including E230M does not significantly alter the optical+IR posteriors and SiO remains only weakly constrained. 

Using forward-model comparisons with the \texttt{sunbather} grid, we find that the broadband E230M spectrum does not meaningfully distinguish among the tested compositions, with small AIC differences and broadly distributed model probabilities within the model set. As a guide for the upper-atmosphere modeling, ATES estimates a mass-loss rate of $\dot{M}\sim10^{12.14}\,\mathrm{g\,s^{-1}}$ under our assumed high-energy irradiation proxy. In contrast, the targeted Mg\,{\sc ii} bins provide substantially stronger discrimination: the extreme case of $10\times$ metallicity with $10\times$ solar Mg abundance is strongly disfavored while models without extreme Mg enhancement remain competitive.  Taken together, the Mg\,{\sc ii} observations favor a scenario in which the measured line-wing effective radii can be produced without requiring unusually large Mg abundances, instead pointing to high-altitude Mg\,{\sc ii} distributed over an extended path length. Together with the near-solar metallicity inferred at lower altitudes \citep{Fu_2021}, the Mg\,{\sc ii} constraints are qualitatively consistent with a hydrodynamic outflow that does not require strong Mg enrichment and is well mixed.

Finally, we emphasize that interpreting Mg\,{\sc ii} in terms of physical extent and mass-loss rate requires sophisticated modeling that couples hydrodynamic structure, ionization balance, and radiative transfer; independent constraints such as those published He\,{\sc i} 10830 {\AA} upper limits \citep{Lampon_2023} provide valuable bounds on $\dot{M}$ for such models.

Looking forward from an observational perspective, expanding the number of UHJs with NUV constraints is essential for placing individual detections in a population context. Ongoing HST efforts (e.g., GO-18108; PI: Dos Santos) will increase the sample of UHJs observed in the UV and enable comparative tests of how irradiation environment and planetary parameters shape the detectability and kinematics of escaping metals.\\ 

\section*{Acknowledgments}
We thank the anonymous referee for their review and constructive comments. This work is based on observations made with the NASA/ ESA Hubble Space Telescope obtained at the Space Telescope Science Institute (STScI), which is operated by the Association of Universities for Research in Astronomy, Inc. The observations associated with this analysis are the HST-GO programs 16724 and the datasets can be accessed via this \dataset[DOI]{https://doi.org/10.17909/4s30-zk02}. Support for this work was provided by NASA through grants under the HST GO-16724 program and HST GO-16634 from STScI. AAV acknowledges funding from the Dutch Research Council (NWO), with project number VI.C.232.041 of the Talent Programme Vici. This project has received funding from the Swiss National Science Foundation for projects 200021\_200726 and 2000-1-243070. It has also been carried out within the framework of the National Centre of Competence in Research PlanetS supported by the Swiss National Science Foundation under grant 51NF40\_205606. This work was supported by grants from eSSENCE (grant number eSSENCE@LU 9:3), the Swedish National Research Council (project number 2023-05307) and the Royal Physiographic Society of Lund, through The Fund of the Walter Gyllenberg Foundation. ML acknowledges support of the Swiss National Science Foundation under grant number PCEFP2\_194576. The contribution of ML has been carried out within the framework of the NCCR PlanetS supported by the Swiss National Science Foundation under grants 51NF40\_182901 and 51NF40\_205606. 

\vspace{5mm}
\facilities{HST(STIS)}

\software{\texttt{astropy} \citep{astropy_2013, astropy_2018}, \texttt{barycorrpy} \citep{Kanodia_2018}, \texttt{batman} \citep{batman}, \texttt{calstis} \citep{calstis}, \texttt{dynesty} \citep{dynesty}, \texttt{george} \citep{george}, \texttt{juliet} \citep{juliet}, \texttt{jupyter} \citep{jupyter}, \texttt{matplotlib} \citep{matplotlib}, \texttt{multinest} \citep{Feroz_2008,Feroz_2009, Feroz_2019}, \texttt{NumPy} \citep{numpy}, \texttt{pandas} \citep{pandas,pandas_2}, \texttt{petitRADTRANS} \citep{Molliere_2019}, \texttt{SciPy} \citep{scipy}, \texttt{stissplice} \citep{stissplice}, \texttt{sunbather} \citep{sunbather}, \texttt{transitspectroscopy} \citep{transitspectroscopy}.}

\appendix \label{Appendix}

\section{Raw Data Visit A and D} \label{Appendix: Raw Visits}

\begin{figure}[h]
    \includegraphics[width=1 \textwidth]{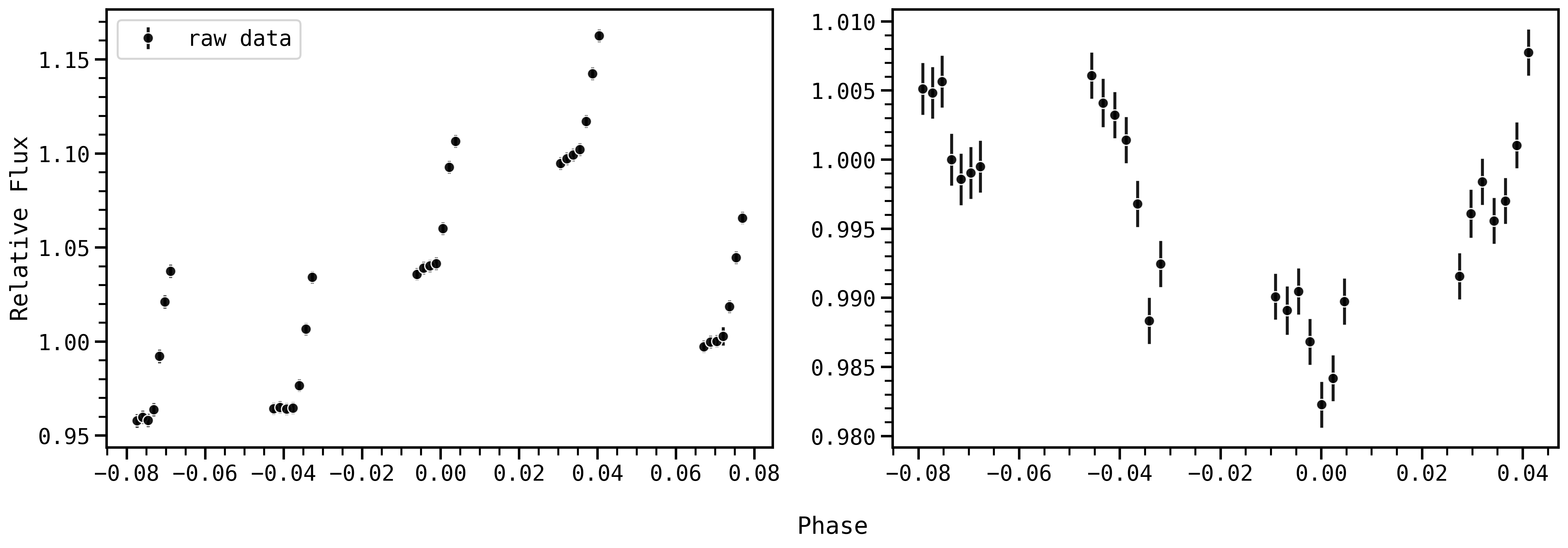}
    \caption{Phase-folded raw white light curves for Visit A (left) and Visit D (right). Normalized with respect to their out-of-transit orbits.}
    \label{fig:10}
\end{figure}

Visit A exhibits substantially larger visit-long and orbit-to-orbit flux variations than visit D (see \autoref{fig:10}), including step-like orbit-to-orbit offsets consistent with changes in slit throughput. Visit A included an ACQ/PEAK \footnote{\url{https://hst-docs.stsci.edu/stisihb/chapter-8-target-acquisition/8-3-onboard-target-acquisition-peakups-acq-peak}} re-centering exposures, whereas visit D used an ACQ sequence without ACQ/PEAK. In particular ACQ/PEAK, re-centers the target in the slit and can introduce discontinuities in the throughput between HST orbits. Because these effects dominate the broadband continuum and are not robustly captured by our systematics model, we exclude visit A from the broadband and narrow-band analyses. We note that the velocity-resolved single line analysis is performed using relative in/out-of-transit spectra and is therefore less sensitive to absolute throughput offsets, allowing us to include visit A for line-profile comparisons. We do not recommend using acquisition strategies that can introduce orbit-to-orbit throughput offsets (e.g., ACQ/PEAK) for STIS/E230M time-series transit spectrophotometry, as these steps can dominate the broadband light curve and complicate systematics correction. 

\section{Jitter Detrending Parameters and Results}

We used the jitter engineering data for our systematics detrending as discussed in \autoref{sec:Systematics}. \autoref{tab: jitter_all} shows a full list of the jitter vectors available to use. In \autoref{tab: jitter_coeff} we can see a summary of the resulting median coefficients of the jitter vectors after the white light curve and spectroscopic fits. 

\begin{table}[h]
    \centering
    \begin{tabular}{c c}
    \hline\hline
    Parameters & Definition\\
    \hline
    \texttt{V2\_dom} & Dominant fine guidance sensor V2 coordinate\\
    \texttt{V3\_dom} & Dominant fine guidance sensor V3 coordinate\\
    \texttt{V2\_roll} & Roll fine guidance sensor V2 coordinate\\ 
    \texttt{V3\_roll} & Roll fine guidance sensor V3 coordinate\\
    \texttt{SI\_V2\_avg} & Mean jitter in V2 over 3 seconds\\
    \texttt{SI\_V2\_rms} & RMS jitter in V2 over 3 seconds\\
    \texttt{SI\_V2\_p2p} & Peak-to-peak jitter in V2 over 3 seconds\\
    \texttt{SI\_V3\_avg} & Mean jitter in V2 over 3 seconds\\
    \texttt{SI\_V3\_rms} & RMS jitter in V2 over 3 seconds\\
    \texttt{SI\_V3\_p2p} & Peak-to-peak jitter in V2 over 3 seconds\\
    \texttt{RA} & Right ascension of aperture reference\\
    \texttt{DEC} & Declination of aperture reference\\
    \texttt{Roll} & Position angle between north and $+$V3\\
    \texttt{Limbang} & Angle between V1 axis and Earth limb\\
    \texttt{Termang} & Angle between V1 axis and terminator\\
    \texttt{LOS\_zenith} & Angle between HST zenith and target\\
    \texttt{Latitude} & HST subpoint latitude\\ 
    \texttt{Longitude} & HST subpoint longitude\\ 
    \texttt{Mag\_V1} & Magnetic field along V1\\
    \texttt{Mag\_V2} & Magnetic field along V2\\
    \texttt{Mag\_V3} & Magnetic field along V3\\
    \hline
    \end{tabular}
    \label{tab: jitter_all}
    \caption{Jitter engineering data optical state vectors. See \url{https://www.stsci.edu/hst/instrumentation/focus-and-pointing/pointing/jitter-file-format-definition} for definitions.}
\end{table}

\begin{table}[h]
    \caption{Median jitter parameter coefficients for the white light curve and the spectroscopic fits.}
    \centering
    \begin{tabular}{c c c c c c c c c c}
    \hline\hline
    & & \texttt{V2\_dom} & \texttt{V2\_roll} & \texttt{V3\_roll} & \texttt{SI\_V2\_avg} & \texttt{RA} & \texttt{DEC} & \texttt{Latitude} & \texttt{Longitude} \\ 
    \hline
    \multirow{2}{*}{White light} &  GP & 0.3855 & 0.5271 & 0.4517 & 0.5191 & 0.3804 & 0.3782 & 0.5635 & 0.4685 \\ 
    & Linear & 0.0010 & 0.0007 & 0.0002 & -0.0013 & -0.0013 & -0.0032 & -0.0001 & -0.0015\\ 
    \hline
    \multirow{2}{*}{Bin 1} &  GP & 0.5671 & 0.7088 & 0.6325 & 0.6911 & 0.4762 & 0.5192 & 0.6927 & 0.6455\\ 
    & Linear & 0.0011 & 0.0026 & 0.0007 & -0.0011 & -0.0004 & -0.0029 & 0.0005 & -0.0020 \\ 
    \hline
    \multirow{2}{*}{Bin 2} &  GP & 0.3972 & 0.7020 & 0.5854 & 0.5513 & 0.4739 & 0.4407 & 0.7320 & 0.5619\\ 
    & Linear & 0.0010 & 0.0023 & 0.0011 & -0.0017 & -0.0003 & -0.0043 & 0.0005 & -0.0008\\ 
    \hline
    \multirow{2}{*}{Bin 3} &  GP & 0.4656 & 0.7216 & 0.5829 & 0.6526 & 0.4720 & 0.4270 & 0.7977 & 0.5658\\ 
    & Linear & 0.0009 & -0.0002 & 0.0004 & -0.0010 & -0.0014 & -0.0039 & 0.0005 & -0.0012 \\ 
    \hline
    \multirow{2}{*}{Bin 4} &  GP & 0.4619 & 0.5888 & 0.6648 & 0.6462 & 0.5330 & 0.2984 & 0.8093 & 0.5805\\ 
    & Linear & 0.0010 & -0.0012 & -0.0003 & -0.0011 & -0.0033 & -0.0025 & -0.0005 & -0.0014\\ 
    \hline
    \end{tabular}
    \label{tab: jitter_coeff}
\end{table}

\section{Additional Spectra from Single Line Analysis: Mg\,{\sc i}, Fe\,{\sc i}, and Fe\,{\sc ii}}

\begin{figure*}[h]
    \includegraphics[width=1 \textwidth]{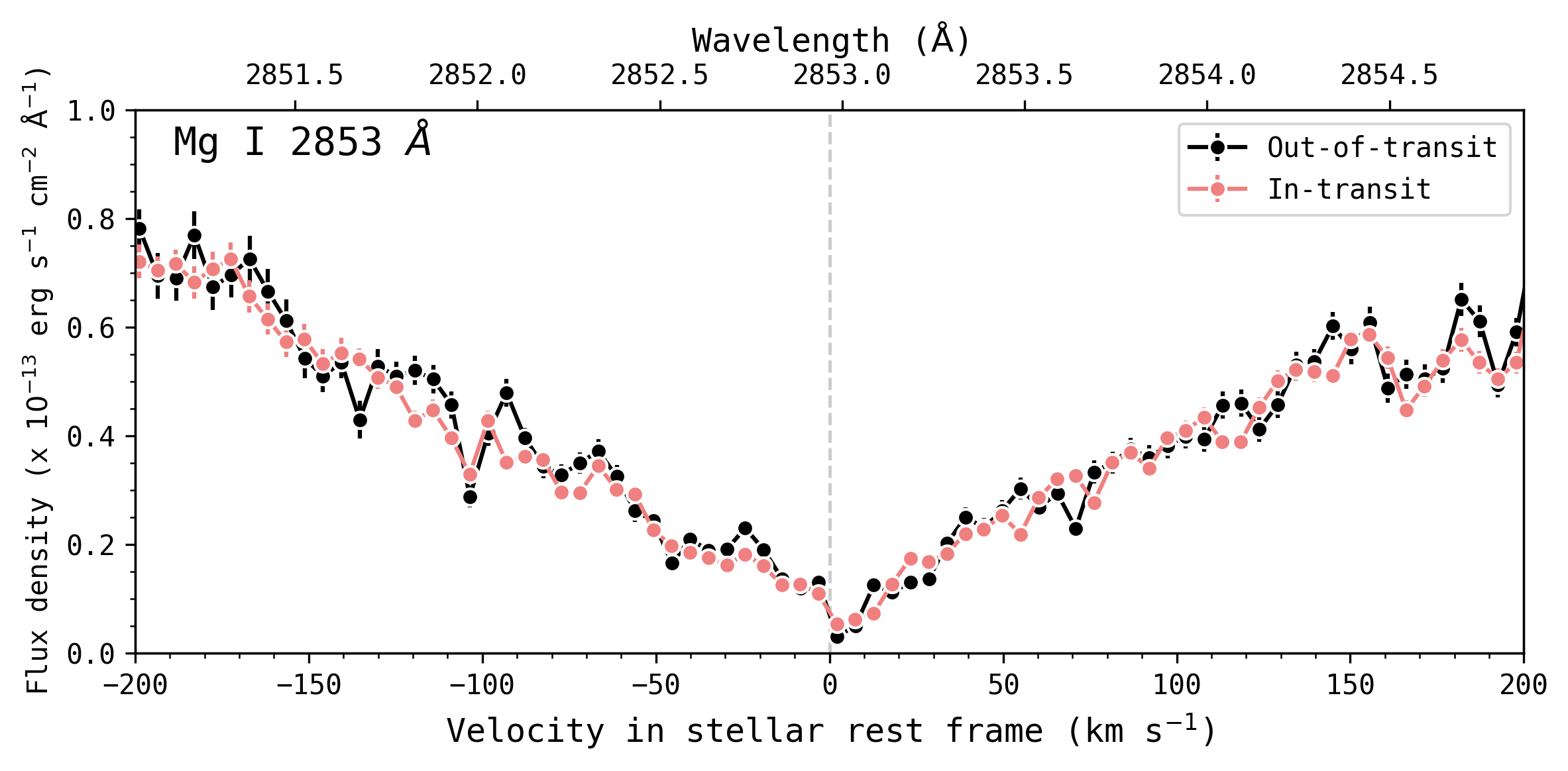}
    \caption{Same as \autoref{fig:5} but for Mg\,{\sc i} at 2852.965 {\AA}. No significant excess absorption is seen in this line.}
    \label{fig:11}
\end{figure*}

\begin{figure*}[h]
    \includegraphics[width=1
    \textwidth]{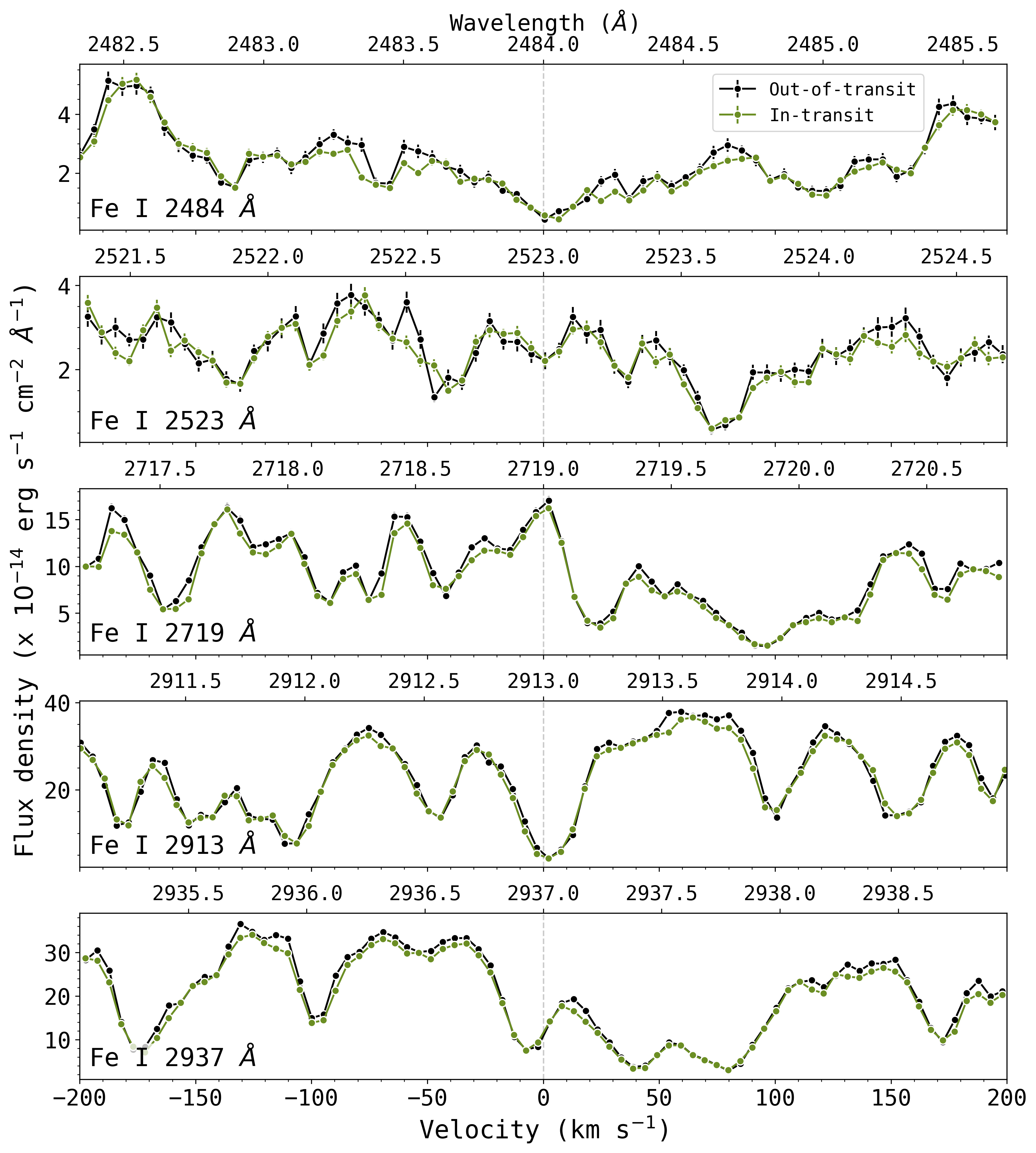}
    \caption{Same as \autoref{fig:5} but for Fe\,{\sc i} at 2484 {\AA}, 2523 {\AA}, 2719 {\AA}, 2913 {\AA} and 2937 {\AA}. No significant excess absorption is seen in these lines.}
    \label{fig:12}
\end{figure*}

\begin{figure*}[h]
    \includegraphics[width=1 \textwidth]{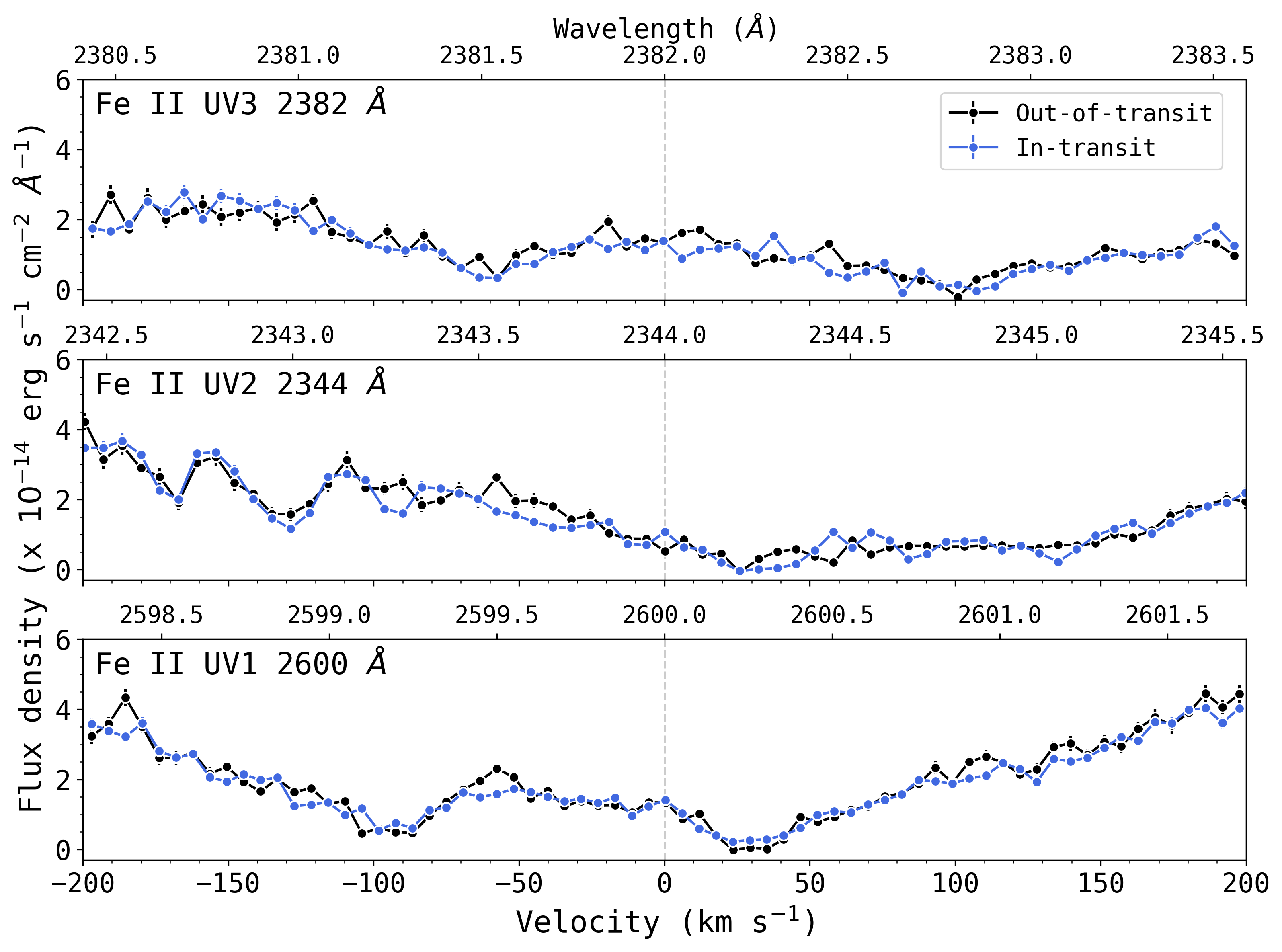}
    \caption{Same as \autoref{fig:5} but for the Fe\,{\sc ii} multiplets at at 2382 {\AA}, 2344 {\AA} and 2600 {\AA}. No significant excess absorption is seen in these lines.}
    \label{fig:13}
\end{figure*}

In \autoref{sec: Results - high res} we discussed our single line analysis for the metal species. \autoref{fig:11} shows our results for Mg\,{\sc i}, \autoref{fig:12} for Fe\,{\sc i} and \autoref{fig:13} for Fe\,{\sc ii}.

\section{Targeted Mg Light Curves from 0.5 {\AA} bins} \label{Appendix: Mg II}

\begin{figure*}[h]
    \includegraphics[width=1 \textwidth]{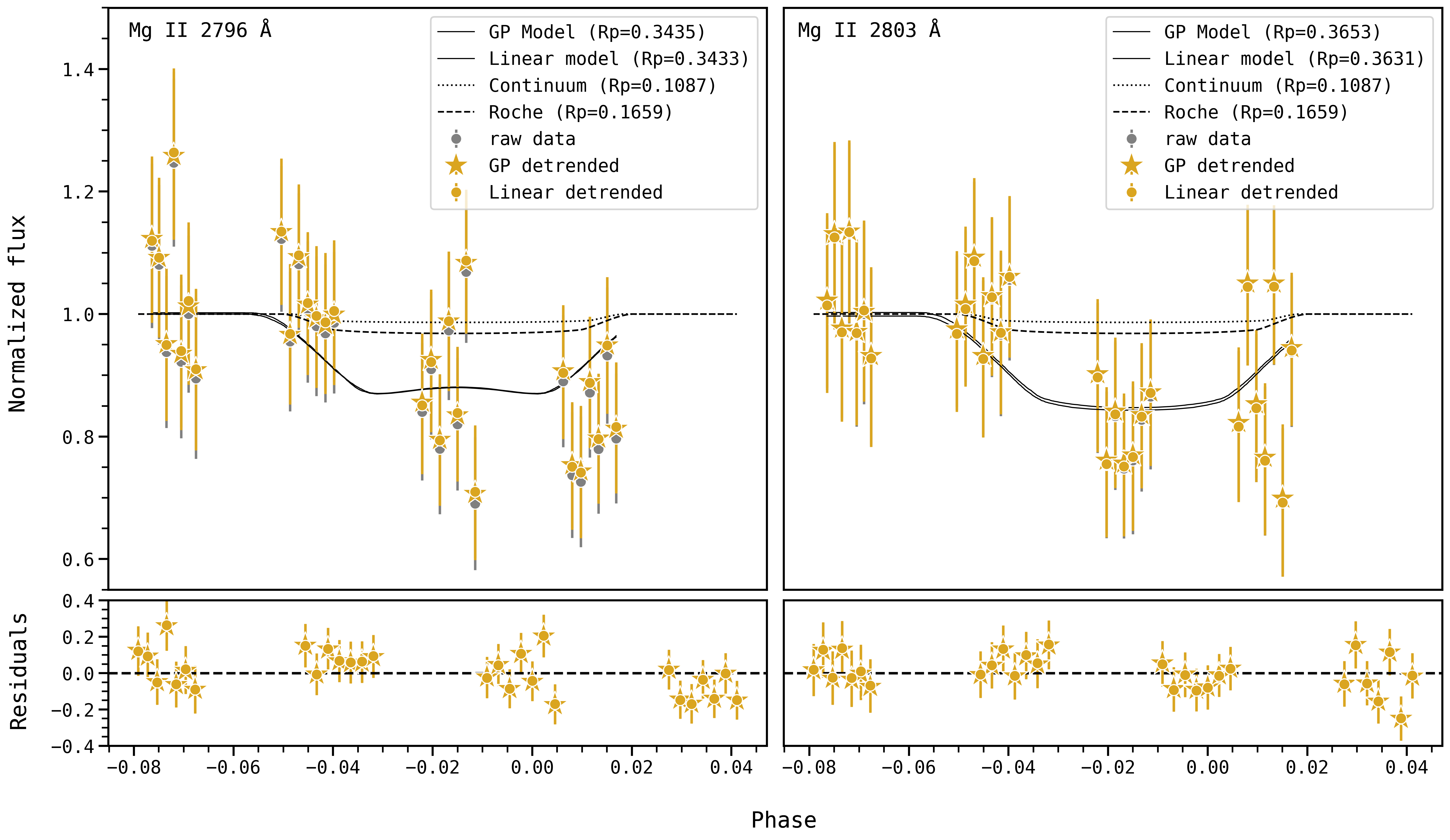}
    \caption{Targeted 0.5 {\AA} Mg\,{\sc ii} light curves for the blue wing bins used in \autoref{fig:7} (2796.1 {\AA} for the $k$ line and 2803.28 {\AA} for the $h$ line). Gray points show the normalized raw time series for visit D while gold points show the corresponding detrended time series using the GP (stars) and linear (circles) systematics models. Solid curves show the best-fit transit models for each detrending treatment. The dotted curve indicates the expected optical transit depth from \citep{Fu_2021} and the dashed curve shows a Roche-lobe occultation benchmark. Bottom panels show residuals relative to the best-fit models.}
    \label{fig:14}
\end{figure*}

\autoref{fig:14} shows the light curves for the two 0.5 {\AA} bins centered on the Mg doublet used to generate the targeted Mg\,{\sc ii} transmission spectrum in \autoref{fig:7} (2796.1 {\AA} and 2803.28 {\AA}). We show the raw normalized time series and the corresponding GP- and linear- detrended versions along with the best-fit transit models. The GP and linear treatments yield consistent best-fit $R_{\rm eff}/R_s$, indicating that the targeted-bin measurements are not strongly sensitive to the choice of systematics model at the achieved precision.  

\section{Model-data comparison} \label{Model_comparison_indiv - stats}

In \autoref{sec: model comparison} we compared our grid of upper-atmosphere forward models to the combined STIS/E230M dataset (four broadband points plus the 22 targeted Mg\,{\sc ii} bins) and found that the Mg\,{\sc ii} measurements dominate the model discrimination. To make this statement explicit we followed the same procedure described in \autoref{sec: model comparison} separately for the E230M broadband subset (N = 4) and for the targeted Mg\,{\sc ii} subset (N = 22), under both systematics treatments (GP and linear). The results are summarized in \autoref{tab: model_comparison_indiv}. For the broadband points alone the discrimination is weak: under the GP treatment the sub‑solar model ($z = 0.1$) has the smallest AIC ($\Delta$AIC $\leq$0.34 across the grid) and models receive broadly similar Akaike weights (each $\approx$ 15–18\%), while under the linear treatment the $z = 10$, Mg$\times$10 model attains the lowest AIC but again the $\Delta$AIC values among models are small ($\leq$1.8), so no single composition is decisively preferred by the broadband data. The targeted Mg\,{\sc ii} bins, in contrast, yield much stronger discrimination: under the GP treatment the solar composition ($z = 1$) attains the minimum AIC and the largest model weight ($p(M)$ $\approx$ 27.5\%), with $z = 0.1$ and $z = 100$ remaining competitive ($\Delta$AIC $\leq$ 0.5); under the linear treatment $z = 1$ again ranks highest ($p(M)$ $\approx$ 28.6\%) with $z = 0.1$ and $z = 100$ close behind ($\Delta$AIC $\leq$ 0.66). In both systematics treatments the model with extreme Mg enhancement at high metallicity ($z = 10$, Mg$\times$10) is strongly disfavored by the Mg\,{\sc ii} bins ($\Delta$AIC $\approx$ 16.3–16.5; $p(M)$ $\leq$ 0.01\%, i.e., $p(\neg M)=1-p(M)$ $\approx$ 99.9 within the tested set). These AIC weights are conditional on the finite model grid and equal prior weights across models, and therefore quantify relative support within this tested model set rather than absolute probabilities. The robust outcome from these per‑dataset comparisons is that (i) the broadband E230M points by themselves do not single out a preferred composition at the present precision, and (ii) the Mg\,{\sc ii} line‑wing measurements strongly disfavor models that overproduce Mg\,{\sc ii} opacity (the $z = 10$, Mg$\times$10 case) while being consistent with a range of moderate metallicities and Mg abundances within the tested grid.

\begin{table*}[t]
\centering
\caption{\texttt{sunbather} model comparison to the STIS/E230M broadband spectrum (N = 4) and the targeted Mg~II 0.5 {\AA} bins (N = 22) for GP and linear systematics treatment. We report $\chi^2$, AIC, $\Delta$AIC (relative to the best model within each subset), and AIC model probabilities $p(M)$ normalized within each subset. We also report $p(\neg M)=1-p(M)$ as the disfavored probability within the tested model set.}
\begin{tabular}{l l r r r r r}
\hline
\hline
Systematics & Model & $\chi^2$ & AIC & $\Delta$AIC & $p(M)$ (\%) & $p(\neg M)$ (\%)\\
\hline
\multicolumn{7}{c}{E230M broadband (N = 4)}\\
\hline
GP     & $z=0.1$ & 0.35 & 0.35 & 0.00 & 17.6 & 82.4\\
GP     & $z=1$ & 0.35 & 0.35 & $2.73\times10^{-4}$ & 17.6 & 82.4\\
GP     & $z=10$ & 0.45 & 0.45 & 0.10 & 16.8 & 83.4\\
GP     & $z=1$, Mg$\times$10 & 0.46 & 0.46 & 0.11 & 16.7 & 83.3\\
GP     & $z=100$ & 0.49 & 0.49 & 0.14 & 16.5 & 83.5\\
GP     & $z=10$, Mg$\times$10 & 0.69 & 0.69 & 0.34 & 14.9 & 85.1\\
\hline
Linear & $z=10$, Mg$\times$10 & 1.66 & 1.66 & 0.00 & 26.3 & 73.7\\
Linear & $z=100$ & 2.36 & 2.36 & 0.70 & 18.6 & 81.4\\
Linear & $z=10$ & 2.49 & 2.49 & 0.83 & 17.4 & 82.6\\
Linear & $z=1$, Mg$\times$10 & 2.69 & 2.69 & 1.03 & 15.7 & 84.3\\
Linear & $z=1$ & 3.40 & 3.40 & 1.74 & 11.0 & 89.0\\
Linear & $z=0.1$ & 3.41 & 3.41 & 1.75 & 11.0 & 89.0\\
\hline
\multicolumn{7}{c}{Targeted Mg\,{\sc ii} 0.5 {\AA} bins (N = 22)}\\
\hline
GP     & $z=1$ & 5.64 & 5.64 & 0.00 & 27.5 & 72.5\\
GP     & $z=0.1$ & 5.70 & 5.70 & 0.05 & 26.8 & 73.2\\
GP     & $z=100$ & 6.13 & 6.13 & 0.49 & 21.6 & 78.4\\
GP     & $z=10$ & 6.59 & 6.59 & 0.95 & 17.1 & 82.9\\
GP     & $z=1$, Mg$\times$10 & 8.40 & 8.40 & 2.75 & 6.95 & 93.0\\
GP     & $z=10$, Mg$\times$10 & 22.1 & 22.1 & 16.5 & 0.01 & 99.9\\
\hline
Linear & $z=1$ & 5.75 & 5.75 & 0.00 & 28.6 & 71.4\\
Linear & $z=0.1$ & 5.80 & 5.80 & 0.05 & 27.9 & 72.1\\
Linear & $z=100$ & 6.41 & 6.41 & 0.66 & 20.6 & 79.4\\
Linear & $z=10$ & 6.87 & 6.87 & 1.12 & 16.3 & 83.7\\
Linear & $z=1$, Mg$\times$10 & 8.67 & 8.67 & 2.91 & 6.65 & 93.4\\
Linear & $z=10$, Mg$\times$10 & 22.1 & 22.1 & 16.3 & 0.01 & 99.9\\
\hline
\end{tabular}
\label{tab: model_comparison_indiv}
\end{table*}

\clearpage
\bibliographystyle{aasjournal}
\bibliography{references}{}

\end{document}